\documentclass[lettersize,journal,article]{IEEEtran}
\usepackage{subcaption}
\usepackage{amsmath,amsfonts}
\usepackage{booktabs}
\usepackage{algorithmic}
\usepackage{algorithm}
\usepackage{amssymb}
\usepackage{array}
\usepackage[caption=false,font=normalsize,labelfont=sf,textfont=sf]{subfig}
\usepackage{textcomp}
\usepackage{stfloats}
\usepackage{url}
\usepackage{verbatim}
\usepackage{graphicx}
\usepackage{cite}
\usepackage{titlesec}
\usepackage{subfig}
\usepackage{CJK}
\usepackage{ragged2e}
\usepackage{lipsum}
\usepackage{xcolor}
\usepackage{tcolorbox}
\usepackage{multicol}
\usepackage{array}
\usepackage{multirow}
\usepackage{float}
\usepackage{cuted}
\usepackage{changepage}
\usepackage{titlesec}
\usepackage{tabularx}
\usepackage{todonotes}

\newcommand{\ie}{i.\,e.,\ }

\begin{document}

% \title{Pioneering the Next-Generation of Emotion Intelligence in the LLM Era: Open Vocabulary Emotion Understanding}
\title{Pioneering Multimodal Emotion Recognition in the Era of Large Models:\\ From Closed Sets to Open Vocabularies}
%BS: introduced a line break for nice appearance :)

% \author{Zixing Zhang,~\IEEEmembership{Senior Member, IEEE,} Zhiqiang Gao, Shihao Gao

%BS: added white space and backslash - will edit for language and minor things directly and leave %BS: comments, otherwise...
\author{Jing Han, Zhiqiang Gao, Shihao Gao, Jialing Liu, Hongyu Chen,\\ Zixing Zhang, Bj\"{o}rn W.\ Schuller

%BS: References incoherently use "et al." - sometimes after a few and sometimes after a few more (see, cf. [9]) authors - I suggest not using et al. for the full references if possible

%BS: In figure 1, "Leat-to-Most" --> "Least-to-Most"  (lower right corner of the image).

\thanks{The work leading to this research was supported by the National Natural Science Foundation of China under Grant No.~U25A20447 and No.~62571184, the Department of Science and Technology of Hunan Province under Grant No.~2025RC6003, the Guangdong Basic and Applied Basic Research Foundation under Grant No.~2024A1515010112, the Changsha Science and Technology Bureau Foundation under Grant No.~kq2402082, and the Shenzhen Natural Science Foundation under Grant No.~JCYJ20250604190534043. (Corresponding authors: Zixing~Zhang)}
\thanks{J.~Han is with the Department of Computer Science and Technology, University of Cambridge, CB3 0FD Cambridge, U.K. (e-mail: jh2298@cam.ac.uk)}
\thanks{Z.~Gao, S.~Gao, J.~Liu, H.~Chen, and Z. Zhang are with the College of Computer Science and Electronic Engineering, Hunan University, Changsha 410082, China. (e-mail: \{gaozhiqiang, shihaogao, ljl1, redtea, zixingzhang\}@hnu.edu.cn)} %jhan, 
\thanks{Z.~Zhang is also with the Shenzhen Research Institute, Hunan University, Shenzhen 518000, China.}
\thanks{Bj\"{o}rn W. Schuller is with GLAM -- the Group on Language, Audio, and  Music, Imperial College London, SW7 2BX London, U.K., and also with CHI -- the Chair of Health Informatics at TUM University Hospital, the MCML -- Munich Center for Machine Learning, and the MDSI -- Munich Data Science Institute, all in Munich, Germany. (e-mail: schuller@ieee.org)}

}

% Todo: change article headers
% \markboth{AIOpen}
% {Shell \MakeLowercase{\textit{et al.}}: A Sample Article Using IEEEtran.cls for IEEE Journals}

% Todo: change article tail
% \IEEEpubid{0000--0000/00\$00.00~\copyright~2021 IEEE}

\maketitle

\begin{abstract}

Recent advances in multimodal large language models (MLLMs) have demonstrated remarkable multi- and cross-modal integration capabilities. However, their potential for fine-grained emotion understanding remains systematically underexplored. While open-vocabulary multimodal emotion recognition (MER-OV) has emerged as a promising direction to overcome the limitations of closed emotion sets, no comprehensive evaluation of MLLMs in this context currently exists. To address this, our work presents the first large-scale benchmarking study of MER-OV on the OV-MERD dataset, evaluating \textit{19 mainstream MLLMs}, including general-purpose, modality-specialized, and reasoning-enhanced architectures. Through systematic analysis of model reasoning capacity, fusion strategies, contextual utilization, and prompt design, we provide key insights into the capabilities and limitations of current MLLMs for MER-OV. 
Our evaluation reveals that a two-stage, trimodal (audio, video, and text) fusion approach achieves optimal performance in MER-OV, with video emerging as the most critical modality. We further identify a surprisingly narrow gap between open- and closed-source LLMs.
These findings establish essential benchmarks and offer practical guidelines for advancing open-vocabulary and fine-grained affective computing, paving the way for more nuanced and interpretable emotion AI systems. Associated code will be made publicly available upon acceptance.
% The experimental code will be available at \url{https://github.com/warmbreeze92/MER-OV-Bench/tree/main}.
\end{abstract}


\section{Introduction}

Emotion, a complex interplay of subjective experience, cognitive processing, and physiological response, is fundamental to human interaction~\cite{han2019adversarial}. The ability to automatically recognize emotions is therefore a critical enabling technology in diverse fields, including human-computer interaction, healthcare, and marketing~\cite{kamble2023comprehensive}. Historically, emotion recognition has been dominated by two main paradigms: discrete models that classify emotions into basic categories like happiness or anger~\cite{ekman1992argument}, and dimensional models that map affective states onto continuous axes such as valence, arousal, and dominance~\cite{mehrabian1996pleasure}. However, these traditional approaches share a fundamental limitation: they constrain the complexity of human emotion to a predefined and closed set of labels or axes. This simplification fails to adequately capture the nuanced, blended, and culturally specific emotional states that characterize the real-world human experience.

% To overcome these limitations, the field is increasingly turning toward \textbf{open-vocabulary multimodal emotion recognition (MER-OV)}~\cite{lian2024mer}. This paradigm shifts away from fixed categories, instead leveraging an unrestricted, natural language lexicon to describe affective states~\cite{wu2024towards}. The advantages of the open-vocabulary approach stem from its alignment with modern psychological and sociological theories. From a psychological perspective, emotions are not discrete islands but complex, continuous, and multi-layered phenomena~\cite{rached2013emotion}; MER-OV accommodates this by enabling for fine-grained descriptions that capture subtle interactions and dynamic transitions between emotions. From a sociological standpoint, emotional expression is deeply shaped by cultural norms and interpersonal context~\cite{bericat2016sociology}. MER-OV supports this by allowing the use of culturally specific terminology, thereby avoiding the simplistic and often biased mapping onto a universal but limited set of categories.
To overcome these limitations, the field is increasingly turning toward \textbf{open-vocabulary emotion recognition}. This paradigm shifts away from fixed categories, instead leveraging an unrestricted, natural language lexicon to describe affective states~\cite{wu2024towards}. The advantages of the open-vocabulary approach stem from its alignment with modern psychological and sociological theories. From a psychological perspective, emotions are not discrete islands but complex, continuous, and multi-layered phenomena~\cite{rached2013emotion}; open-vocabulary emotion recognition accommodates this by enabling for fine-grained descriptions that capture subtle interactions and dynamic transitions between emotions. From a sociological standpoint, emotional expression is deeply shaped by cultural norms and interpersonal context~\cite{bericat2016sociology}. Open-vocabulary supports this by allowing the use of culturally specific terminology, thereby avoiding the simplistic and often biased mapping onto a universal but limited set of categories.
% \jh{i somehow felt OV is the opposite of close set as mentioned above, but here it mixed/combined two concepts, OV and multimodal. Would it better to seperately describe these two? OV is the  key one, multimodal is only need to justified a bit that the trend is to leverage information from multiple sources? which is quite norm already?}

The recent rapid development of large language models (LLMs), including their multimodal variants such as Video-LLMs and Audio-LLMs, has catalyzed a paradigm shift in artificial intelligence. The transition from task-specific discriminative models to prompt-driven generative models has unlocked unprecedented capabilities in nuanced language understanding and generation~\cite{Pre-Trained_Language_Models}. This breakthrough is particularly consequential for \textbf{open-vocabulary multimodal emotion recognition (MER-OV)}, as it provides tools that can generate the rich, fine-grained, and context-aware emotional descriptions required by the open-vocabulary approach, thereby advancing it from a theoretical ideal to a practical reality. However, MER-OV is still in its infancy, having been first formalized by the ACM Multimedia 2024 challenge~\cite{lian2024mer}, motivating this in-depth investigation. The scarcity of subsequent research motivates this in-depth investigation.

Building upon the emotional clue-based two-stage method from the challenge~\cite{zhang2024open,Multimodal_Emotion_Captioning}, we propose and evaluate two novel architectures: an objective description-based two-stage method and a video-LLM-only one-stage method. Moreover, we conduct a systematic analysis of unimodal, bimodal, and trimodal fusion strategies to precisely quantify the contribution of each modality and identify optimal combinations. In addition, we provide a broad comparative analysis of leading open-source and closed-source models (LLMs, Video-LLMs, and Audio-LLMs), establishing a comprehensive performance benchmark for the MER-OV task. Given the potential significance of the visual stream, we further perform a targeted investigation into the video modality, including exploration of various frame sampling strategies and novel methods for leveraging video metadata. Additionally, we undertake a comprehensive examination of prompt engineering, meticulously benchmarking hand-crafted ``hard prompts'' against general-purpose methods, advanced optimization techniques, and the direct application of reasoning models.

In summary, our primary contributions are threefold:
\begin{itemize}
    \item We establish the \textbf{first comprehensive benchmark for MER-OV}, extensively evaluating a wide range of foundational frameworks, LLMs (text, video, audio), and modality fusion strategies.
    \item We present \textbf{an in-depth analysis} that delivers key insights, such as quantifying video as the pivotal modality and revealing the narrow performance gap between open- and closed-source LLMs.
    \item We provide \textbf{a practical guideline} established through systematic experiments on prompting strategies and architectural designs, providing a foundation for future research.
\end{itemize}

\section{Related Work}

% In this section, we review related works in three primary areas: multimodal emotion recognition, Multimodal Large Language Models (MLLMs), and prompt engineering.

\subsection{Multimodal Emotion Recognition}

Research in emotion recognition is typically categorized into unimodal and multimodal paradigms. Unimodal approaches, which analyze affective states from a single data stream, such as text~\cite{10542112,zhang2024re}, audio~\cite{10767298, HanZPS21}, or facial expressions~\cite{canal2022survey,10902013}, often suffer from information loss and susceptibility to noise. To overcome these limitations, multimodal emotion recognition (MER) integrates complementary information from multiple data streams, such as audio-visual~\cite{HanZPS21}, speech-text~\cite{zhao2023speaker}, and audio-visual-text fusions~\cite{zhang2024open,han2019emobed}. The primary objective of traditional MER has been to map these fused features onto predefined, closed-vocabulary labels.%, either as discrete categories (e.\,g., anger) or dimensional values (e.\,g., valence/arousal/dominance)~\cite{Tsai_Bai_Liang_Kolter_Morency_Salakhutdinov_2019,han2018towards, han2017strength}.

% A recent paradigm shift has given rise to MER-OV. 
% Unlike traditional MER, MER-OV requires generating unconstrained, descriptive natural language to represent emotional states. 

% This open-ended formulation renders conventional classification-based methods inapplicable and necessitates the use of LLMs.
% \jh{MER-OV, however, represents a paradigm shift from this closed-set classification to an open-ended generative task. Here, the objective is to produce flexible, descriptive natural language that captures the full spectrum of emotional states.}
% \jh{Consequently, this shift demands models with strong generative capabilities, positioning LLMs as the critical architectural component for tackling the task.}
MER-OV, however, represents a paradigm shift from this closed-set classification to an open-ended generative task. Here, the objective is to produce flexible, descriptive natural language that captures the full spectrum of emotional states.
Consequently, this shift demands models with strong generative capabilities, positioning LLMs as the critical architectural component for tackling the task.
Pioneering work in this area, such as AffectGPT~\cite{lian2025affectgptnewdatasetmodel}, has begun to explore this direction by employing pre-fusion operations to enhance multimodal integration before leveraging the generative power of an LLM. A key challenge in this emerging field is the availability of suitable datasets. While numerous high-quality datasets exist for traditional MER (e.\,g., IEMOCAP~\cite{busso2008iemocap}, CMU-MOSI~\cite{Zadeh2016MOSIMC}, CMU-MOSEI~\cite{zadeh2018multimodal}, MELD~\cite{poria2018meld}), they are universally annotated with closed-vocabulary labels. Consequently, the only publicly available benchmark for MER-OV remains the dataset released for the aforementioned ACM Multimedia 2024 challenge, which we refer to as OV-MERD~\cite{lian2024mer}.

\subsection{Large Language Models and Multimodal LLMs} 

The advent of LLMs, such as the GPT series has marked a pivotal paradigm shift in artificial intelligence~\cite{brown2020language,10447044}. Characterized by their scale and training on vast text corpora, LLMs are distinguished from prior Pre-trained Language Models by their emergent abilities. This capacity for nuanced language understanding and generation is what makes LLMs a foundational technology for open-vocabulary tasks.

% Human emotion, in particular, is a quintessential multimodal phenomenon, expressed through a rich interplay of facial expressions, vocal tones, and linguistic content. This limitation necessitates the development of \textbf{MLLMs}. 
Multimodal LLMs (MLLMs) are designed to process and integrate diverse data types, such as text, images, audio, and video. They have developed rapidly in recent years, especially following the release of GPT-4\cite{achiam2023gpt,3462244}, which sparked widespread research interest due to its impressive multimodal capabilities. Typically, MLLMs augment a frozen LLM backbone with specialized encoders for other modalities, such as vision or audio. A lightweight adapter module, like a perception projector, is then used to map the features from these encoders into the LLM's semantic space, enabling true cross-modal understanding.

Furthermore, the field continues to evolve, with a recent focus on enhancing the multi-step, analytical reasoning capabilities of these models. Models specifically optimized for deep reasoning, which we term Reasoning Models in this article, represent the current state-of-the-art. This evolution directly motivates our evaluation of Reasoning Models within the MER-OV paradigm, to assess whether their advanced inferential capabilities translate to finer-grained emotional understanding.

\subsection{Prompt Engineering}
A prompt is a crucial input that guides a generative model's output~\cite{mesko2023prompt,heston2023prompt}, and empirical evidence confirms that high-quality prompts significantly improve performance~\cite{liu2023pre}. The methodologies for designing these prompts have evolved into a sophisticated hierarchy, which we adopt for our investigation.

\textit{Foundational hard prompts:} At the fundamental level are hard prompts: manually crafted instructions and examples. This category includes foundational In-Context Learning (ICL) techniques~\cite{brown2020language}. The simplest forms are \textbf{Zero-Shot Prompting}, which relies entirely on the model's pre-existing knowledge without any examples, and \textbf{Few-Shot Prompting}, which provides a few input-output exemplars to guide the model. Within this foundational group, we also include \textbf{Chain-of-Thought (CoT)} prompting~\cite{wei2022chain}. CoT enhances reasoning by instructing the model to generate a sequence of explicit, intermediate steps prior to the final answer.
% before arriving at a final answer.

\textit{Composite prompting strategies:} Building upon these foundational prompts, \textbf{composite strategies} integrate more advanced optimization techniques to enhance robustness and accuracy. For instance, \textbf{Self-Consistency}~\cite{wang2022self} improves upon CoT by generating multiple diverse reasoning paths and selecting the final answer via a majority vote. \textbf{Self-Refine}~\cite{madaan2023self} employs an iterative framework where the model generates an initial solution, critiques its own response, and then improves it based on that feedback. Similarly, \textbf{Least-to-Most Prompting}~\cite{zhou2022least} tackles complex problems by first decomposing them into simpler subproblems and then solving them sequentially.
The broad spectrum of prompting strategies, from foundational to composite, defines a vast design space. In this paper, we conduct systematic evaluation of these strategies to establish empirical guidelines for their application in MER-OV.

\begin{figure*}[htbp]
    \centering
    \includegraphics[trim={0cm 0cm 0cm 0cm}, clip, width=\textwidth]{Figures/overview.pdf}
    % \caption{Overview of the proposed multimodal emotion recognition framework.
    % In the first stage, video and audio streams are processed independently by the Video-LLM and Audio-LLM, respectively, using tailored prompts to extract emotion-related clues from each modality. These extracted emotional clues, along with subtitle content, are then integrated into a structured prompt template to construct a comprehensive multimodal scene description. This is combined with various hard prompt designs, forming the input instructions for the second stage.
    % In the second stage, multiple prompt-based optimization strategies, such as integration, iteration, and decomposition, are applied to enhance the LLM’s ability to extract open-vocabulary emotional states. 
    % % Specifically: 
    % % (a) the base prompt is directly input into the LLM to generate initial predictions; 
    % % (b) the integration strategy generates multiple candidate responses, from which the optimal response is selected; 
    % % (c) the iteration strategy introduces a feedback-refine loop, which iteratively revises the output based on model-generated feedback until convergence or a preset threshold is reached; 
    % % (d) the decomposition strategy partitions the original prompt into sub-tasks, each addressed individually by the LLM before being aggregated into the final response.
    % The final output is a list of open-set emotional states associated with the characters in the input video (e.g., concerned, nervous, worried).}
    \caption{Overview of the open-vocabulary multimodal emotion recognition framework based on large language models (LLMs).
    In Stage 1, video and audio streams are processed independently by the Video-LLM and Audio-LLM, respectively, using tailored prompts to extract emotion-related clues from each modality. 
    These extracted emotional clues, along with subtitle content, are then integrated into a structured prompt template to construct a comprehensive multimodal scene description. 
    This is combined with various hard prompt designs, forming the input instructions for the second stage.
    In Stage 2, by leveraging the multimodal scene description, multiple prompt-based optimization strategies, such as integration, iteration, and decomposition, are applied to enhance the LLM’s ability to extract open-vocabulary emotional states. 
    The final output is a list of open-set emotional states associated with the characters in the input video.
    % \jh{this caption needs heavily improved, issues: 1) too long, 2) inconsitent with the figure, e.g., Stage1 vs Stage 1, Video LLM vs Video-LLM, etc, 3) the figure also needs to be improved for better clarity.}
    % \textcolor{blue}{gzq:This figure has been updated. The new caption: Overview of the open-vocabulary multimodal emotion recognition framework based on LLMs. In Stage 1, video and audio streams are independently processed by the Video-LLM and Audio-LLM using tailored prompts to extract emotion-related clues, which are then fused with subtitle content in a structured prompt template to form a comprehensive multimodal scene description. This description, combined with hard prompt designs, serves as the input to Stage 2, where prompt-based optimization strategies (integration, iteration, and decomposition) are applied to enhance the LLM’s ability to extract open-vocabulary emotional states. The framework outputs a list of open-vocabulary emotional states associated with the characters in the input video.}
    }

    \label{overview}
\end{figure*}
% \section{Methodology}
\section{Open-vocabulary multimodal emotion recognition }
% We structure our investigation around four key research questions, each addressed in a dedicated subsection: (1) What is the optimal architectural framework and modality combination? (2) Which models perform best for each component of the framework? (3) How can the contribution of the most critical modality be enhanced? (4) How do different prompting strategies influence the final outcome?
% This section outlines our proposed research questions for the MER-OV, highlighting how each research question to influence the task.
This section introduces the methods in details for answering the raised key questions in MER-OV.

\subsection{Architecture Overview}

% In this section, we provide an overview of the methodological framework used in our MER-OV benchmark study. The input to the system is a video clip, which inherently includes audio and subtitle modalities, enabling a comprehensive multimodal analysis. We propose three architectural variants to address the MER-OV task: (a) emotional clue-based two-stage method, (b) objective description-based two-stage method, and (c) Video-LLM-only one-stage method. Figure~\ref{overview} depicts the emotional clue-based two-stage method as the foundational framework. This architecture is divided into two stages: Multimodal Emotional Clues Generation (Stage 1) and Open-Vocabulary Emotion Inferencing (Stage 2).\jh{it mixed the description of 3 variants with the example overview, I recommend to split to make it clear, ie focus on using one example to describe the two stage, instead of merging the intro of 3 variants, leave that to next subsection. Check if the next paragraph can be used to replace this paragraph.}

This section provides a high-level overview of our MER-OV benchmarking framework, which takes a video clip (with inherent audio and subtitles) as input. Our core architecture is exemplified by the emotional clue-based two-stage method illustrated in Figure~\ref{overview}. Its pipeline is divided into Multimodal Emotional Clues Generation (Stage 1) and Open-Vocabulary Emotion Inferencing (Stage 2). The complete set of three architectural variants is detailed in Section~\ref{sec:method_framework_modality}.

% In Stage 1, the video and audio streams are processed independently by a Video-LLM and an Audio-LLM, respectively, using tailored prompts to extract emotion-relevant clues from each modality (Section~\ref{sec:method_model_benchmark}). These extracted clues, along with subtitle content, are then fused into a structured prompt template. Additional contextual information from the video modality, such as metadata or varied frame sampling outputs, can be incorporated into this template to enrich the multimodal representation (Section~\ref{sec:method_video_enhancement}). The output of Stage 1 is a cohesive multimodal emotional clue.

% In Stage 2, these multimodal clues is combined with various prompt engineering techniques, including integration, iteration, and decomposition strategies, to optimize the LLM's ability to infer open-vocabulary emotional states (Section~\ref{sec:method_prompting}). The final output consists of a set of emotion labels associated with the characters in the input video.
% This overview highlights the modular and extensible nature of our approach, allowing for systematic evaluation across models, fusion strategies, video processing techniques, and prompt designs, as benchmarked in the following sections.

% \jh{Again, the refering to following subsections and the overview is mixed, recommend to seperate as below, pls check carefully, as is created by DS:}

\textit{In Stage 1}, the video and audio streams are processed independently by a Video-LLM and an Audio-LLM, respectively, using tailored prompts to extract emotion-relevant clues from each modality. These extracted clues, along with the subtitle content, are fused into a structured prompt template. Additional contextual information from the video modality, such as metadata or varied frame sampling outputs, can be incorporated to enrich the multimodal representation. The output of Stage 1 is a cohesive set of multimodal emotional clues.

\textit{In Stage 2}, this collection of multimodal clues is combined with various prompt engineering techniques, including integration, iteration, and decomposition strategies, to optimize the LLM's ability to infer open-vocabulary emotional states. The final output is a set of descriptive emotion labels associated with the characters in the input video.

This overview highlights the modular and extensible nature of our approach. We now detail the three architectural variants (Section~\ref{sec:method_framework_modality}), benchmarked models (Section~\ref{sec:method_model_benchmark}), video enhancement techniques (Section~\ref{sec:method_video_enhancement}), and prompting strategies (Section~\ref{sec:method_prompting}) in the following sections.

% the overall architecture integrates the four key research aspects of this paper: (1) method frameworks and multimodal information fusion strategies, (2) the performance of mainstream LLMs and MLLMs on MER-OV with a constructed benchmark, (3) incorporation of additional contextual information from the video modality through techniques such as varying frame sampling strategies and video metadata integration, and (4) application of diverse prompt engineering methods to MER-OV. 

\subsection{Proposed Methods and Multimodal Fusion Strategies}
\label{sec:method_framework_modality}

\begin{figure*}[htbp]
    \centering
    \includegraphics[width=\textwidth]{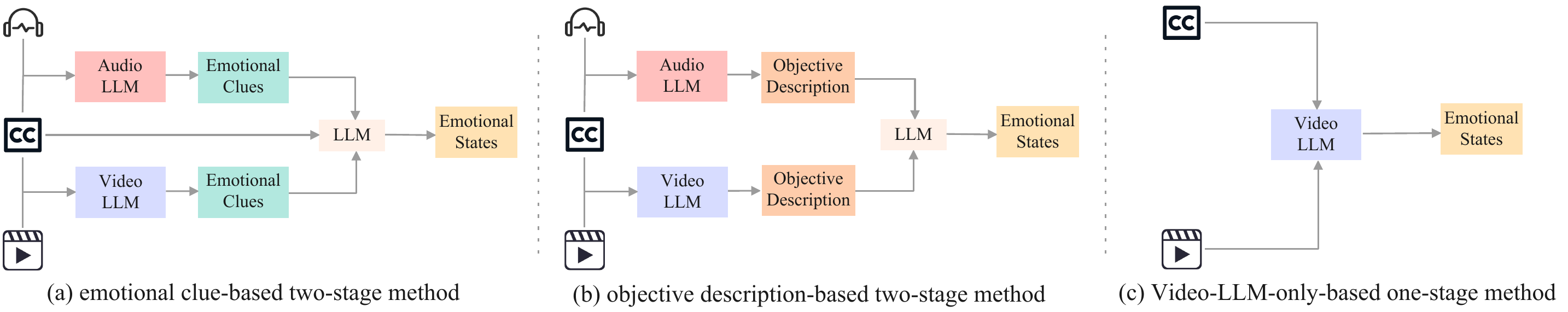}
    \caption{Three LLM-based approaches for open-vocabulary multimodal emotion recognition: (a) emotional clue-based two-stage method; (b) objective description-based two-stage method; and (c) Video-LLM-only-based one-stage method.
    % \jh{what is the line that connected AudioLLM and Video LLM for?}\jh{needs to be improved.}\textcolor{blue}{gzq: This figure has been updated.}
    }
    \label{Figure 1}
\end{figure*}

\begin{figure*}[htbp]
    \centering
    \includegraphics[width=\textwidth]{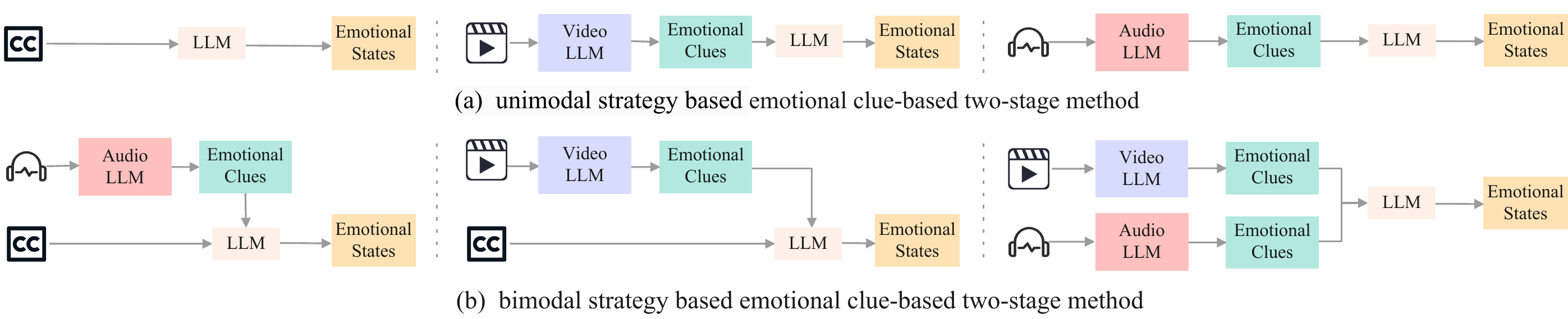}
    \caption{Different modality fusion strategies based on the emotional clue-based two-stage method: (a) unimodal: text, video, and audio; (b) bimodal: video+audio, text+video, and text+audio; (c) trimodal: illustrated in Figure \ref{Figure 1}(a).
    % \jh{it is very rare to see refer to another figure in one figure's caption, pls discuss with Zixing about how to change this}.\jh{need to be improved}
    }
    \label{Figure 2}
\end{figure*}

To establish a robust foundation for MER-OV, we first explore the optimal architectural framework. We propose and compare three distinct methods, as illustrated in Figure~\ref{Figure 1}.

\textit{Emotional Clue-based Two-Stage Method:} It first uses specialized models (an Audio-LLM and a Video-LLM) to extract descriptive \textit{emotional clues} from each modality (e.\,g., tone, pitch for audio; expressions, gestures for video). Subsequently, an LLM integrates these clues with the text subtitles to infer the final emotional state~\cite{lian2023explainable}.
    
\textit{Objective Description-based Two-Stage Method:} In contrast to extracting emotional clues, this method uses the Audio-LLM and Video-LLM to generate detailed but \textit{objective descriptions} of the content within each modality. An LLM then infers based on these neutral descriptions and subtitles.
    
\textit{Video-LLM-only One-Stage Method:} This method investigates a simplified, end-to-end approach. It directly feeds the video input and subtitles into a single Video-LLM, tasking it with jointly analyzing visual and textual information to infer the emotional state.

Having established the architectural framework, we next investigate the relative importance of different modalities. To ensure a controlled comparison, these experiments are conducted within the two-stage framework, using the Emotional Clue-based Two-Stage Method as the benchmark. We systematically evaluate the performance of unimodal (i.\,e., text-only, audio-only, and video-only) and bimodal (i.\,e., text+audio, text+video, audio+video) as illustrated in Figure~\ref{Figure 2}, as well as trimodal 
%BS: added:
approaches 
as depicted in Figure~\ref{Figure 1}. % 

% for the trimodal approach and Figure~\ref{Figure 1} for unimodal and bimodal methods.

\subsection{Model Selection and Evaluation Framework}
\label{sec:method_model_benchmark}

The performance of our two-stage framework is contingent on the individual capabilities of its components: the Audio-LLM, the Video-LLM, and the LLM. As detailed in Figure~\ref{overview}, the process involves:

\textbf{Stage 1.} A Video-LLM and an Audio-LLM extract emotional clues from their respective modalities.

\textbf{Stage 2.} An LLM infers the final emotional state from the generated clues and the source subtitle.

To identify the optimal models for each role, we evaluate a diverse set of seven mainstream LLMs, five Video-LLMs, five Audio-LLMs, and two Reasoning LLMs in Table~\ref{table 1}. This large-scale comparison provides a practical reference for their efficacy on MER-OV.

% \subsection{In-Depth Exploration of the Video Modality}
\subsection{Video Modality Processing Techniques}
\label{sec:method_video_enhancement}

Recognizing that the visual channel contains rich and dense information, we investigate two distinct strategies to enhance its contribution to the final inference. 

\textbf{Frame Sampling Strategies.} The representation of video content for a Video-LLM depends heavily on the frame sampling process~\cite{hu2025m,nie2024slowfocus}.
% \jh{any cite here?}\textcolor{blue}{gzq: add two cite}. 
We compare a \textit{fixed-frame} strategy (24 frames per video) against a \textit{dynamic} strategy. For the dynamic strategy, we set multiple sampling rates: 1, 2, 4, and 6 frames per second.

\textbf{Video Metadata-Augmented.} Beyond visual frames, contextual metadata can provide crucial background information. We investigate a novel method for incorporating video metadata (the video's title and character profiles) as supplementary input to the LLM during the inference stage, to enrich the model's understanding and improve its reasoning process.

\subsection{Prompt Engineering Methods}
\label{sec:method_prompting}

\begin{table*}[t]
\centering
% \caption{Video-LLMs, Audio-LLMs, and LLMs: Architecture, Training Strategy, Training Data Scale, Training Data Composition, Open-Source or Closed-Source, and In.}
\caption{
% Selected recently released five Video-LLMs, five Audio-LLMs, seven LLMs (w/o reasoning), and two LLMs (w/ reasoning), as well as their detailed information, e.g.\jh{e.g. or i.e., be clear when to use which}, backbone architecture, training strategy, training data composition, open/closed-source, and inference ways.\jh{needs to improve the caption and the table format, such as textbf the first line! I give one example here how to improve the caption of the table, please compare the difference and see if it is better: A comprehensive set of models selected for benchmarking, detailing their core attributes. The collection includes \textit{five Video-LLMs}, \textit{five Audio-LLMs}, \textit{seven general-purpose LLMs}, and \textit{two reasoning-enhanced LLMs}, with specifications on aspects such as \textit{backbone architecture}, \textit{training strategy}, \textit{training data composition}, \textit{accessability}, and \textit{inference mode}.}
A comprehensive set of models selected for benchmarking, detailing their core attributes. The collection includes \textit{five Video-LLMs}, \textit{five Audio-LLMs}, \textit{seven general-purpose LLMs}, and \textit{two reasoning-enhanced LLMs}, with specifications on aspects such as \textit{backbone architecture}, \textit{training strategy}, \textit{training data composition}, \textit{accessability}, and \textit{inference mode}.
}
% \zix{is this table introduced in the context?}
\label{table 1}
\begin{tabular}{cccccccc}
\toprule
Modality  & Model   &  Backbone   & Training & Training Data Composition & Access & Inference \\
\midrule
\multirow{5}{*}{\textbf{Video-LLM}} & InternVL2.5-26B~\cite{chen2024expanding} & InternLM2.5-20B        & SFT          & Images/Multi Images/Videos    & Open     & On-Premises \\

& LLaVA-NeXT-Video-7B-DPO                                                     & Vicuna 7B-V1.5         & SFT+DPO      & Images/Video                  & Open     & On-Premises \\

& LLaVA-Video-7B-Qwen2~\cite{zhang2024videoinstructiontuningsynthetic}         & Qwen2-7B               & SFT          & Images/Multi Images/Videos    & Open     & On-Premises \\

& Tarsier2-7B~\cite{yuan2025tarsier2}                                          & Qwen2-7B               & SFT+DPO      & Videos                        & Open     & On-Premises\\

& GPT-4o-mini~\cite{hurst2024gpt}                                              & /                      & /            & Text/Images/Audio/Video       & Closed   & API \\

\midrule
\multirow{5}{*}{\textbf{Audio-LLM}} & Qwen-Audio~\cite{chu2023qwen}            & Qwen-7B                & SFT          & Audio                         & Open     & On-Premises \\

& Qwen2-Audio~\cite{chu2024qwen2}                                              & Qwen2-7B               & SFT+DPO      & Audio                         &  Open    & On-Premises \\

& Gemini1.5-pro~\cite{team2023gemini}                                          & /                      & SFT+RLHF     & Text/Images/Audio/Video       &  Closed & API \\

& Gemini2.0-flash~\cite{Gemini2.0}                                             & /                      & SFT+RLHF     & Text/Images/Audio/Video       & Closed & API \\

& Gemini2.5-pro~\cite{Gemini2.5}                                               & /                       & SFT+RLHF+CoT & Text/Images/Audio/Video       & Closed & API \\

\midrule
\multirow{7}{*}{\textbf{LLM}}       & Gemma2-9B~\cite{team2024gemma}           & Gemma2                  & SFT+RLHF     & Text                          &  Open  & On-Premises \\

\multirow{7.3}{*}{\textbf{(w/o reasoning)}} & Llama3.1-8B~\cite{achiam2023gpt} & Llama3.1               & SFT+RS+DPO   & Text                          &  Open  & On-Premises \\

& Qwen2.5-7B~\cite{yang2024qwen2}                                              & Qwen2.5                & SFT+GRPO     & Text                          &  Open  & On-Premises \\

& Qwen2.5-32B~\cite{yang2024qwen2}                                             & Qwen2.5                & SFT+GRPO     & Text                          &  Open  & On-Premises \\

& Qwen2-72B~\cite{yang2024qwen2technicalreport}                                & Qwen2                  & SFT+RLHF     & Text                          &  Open  & API \\

& DeepSeek-V3~\cite{liu2024deepseek}                                           & DeepSeek-V3            & SFT+GRPO     & Text                          &  Open  & API \\

& GPT-4o-mini                                                                 & /                      & /            & Text/Images/Audio/Video       & Closed  & API \\

\midrule

\multirow{1}{*}{\textbf{LLM}}       & OpenAI o3-mini~\cite{OpenAIo3-mini}      & /                      & SFT+RLHF     & Text                          &  Closed & API \\

\multirow{1.3}{*}{\textbf{(w/ reasoning)}}  & DeepSeek-R1~\cite{guo2025deepseek} & DeepSeek-V3          & SFT+RLHF     & Text                          &  Open    & API \\

\bottomrule
\end{tabular}
\end{table*}

The final stage of our methodology addresses the critical role of prompt engineering in guiding the LLM's inference process. We designed a \textbf{three-stage experimental framework} to systematically evaluate the impact of various prompting techniques, progressing from foundational methods to cutting-edge automated approaches.
% \begin{enumerate}
%     \item We first analyze five fundamental \textbf{hard prompt} designs, including Zero-shot, Few-shot, and CoT patterns.
%     \item We then construct \textbf{composite prompting strategies} by integrating optimization techniques like Self-Consistency and Self-Refine.
%     % \item Next, we explore \textbf{meta-prompting}, a sophisticated APE approach for automated prompt generation.
%     \item Finally, we benchmark these prompting methods against the direct application of advanced \textbf{Reasoning Models}.
% \end{enumerate}
\textbf{Stage 1.} We analyze five fundamental \textbf{hard prompt} designs, including Standard Baseline, Zero-shot, Few-shot, CoT patterns, and Multipersona:
\begin{itemize}
    \item \textbf{Standard Baseline (STD)}: A prompt that only contains a basic task description, similar to a problem statement (without any guiding or prompting information).
    \item \textbf{Zero-shot Chain-of-Thought (Zero-shot-CoT)}: An extension of STD by appending the step-by-step reasoning trigger ``Let's think step by step.'' to guide structured problem-solving.
    \item \textbf{Handcrafted Zero-shot}: A manually designed prompt that incorporates domain-specific heuristics, utilizing common prompt-writing techniques, such as specifying roles, guiding attention to key points, and providing simple processing methods.
    \item \textbf{Handcrafted Few-shot}: A one-shot prompt that uses synthetic examples (due to the lack of in-task training data) to simulate plausible input-output pairs, combined with handcrafted instructions.
    \item \textbf{Multipersona}: Directs the LLM to simulate multiple ``experts'', analyzing the problem from distinct expert perspectives and offering suggestions, ultimately synthesizing a final answer. This process is completed in a single round of interaction.
\end{itemize}

% We then construct \textbf{composite prompting strategies} by integrating optimization techniques like Self-Consistency and Self-Refine. 
% We then construct \textbf{composite prompting strategies} by integrating optimization techniques. To streamline the experimental complexity, we selected one commonly used representative method from each strategy category:

% \begin{itemize}
%     \item \textbf{Ensembling}: Universal Self-Consistency
%     \item \textbf{Self-Criticism}: Self-Refine
%     \item \textbf{Decomposition}: Least-to-Most
% \end{itemize}

\textbf{Stage 2.} We construct \textbf{composite prompting strategies} by integrating optimization techniques. To reduce experimental complexity, we select a representative method from each category: \textbf{Universal Self-Consistency} for ensembling, \textbf{Self-Refine} for self-criticism, and \textbf{Least-to-Most} for decomposition.

\textbf{Stage 3.} We benchmark the STD against the direct application of advanced \textbf{Reasoning Models}. We selected two prominent Reasoning Models (OpenAI o3-mini and DeepSeek-R1)
 to compare with LLMs
 %BS: LLM --> LLMs...
 (GPT-4o-mini and DeepSeek-V3).% To ensure the reliability of these comparisons, all prompt-related evaluations were conducted using a consistent and high-performing set of models identified in our earlier experiments.

\section{Experimental Setup}

\subsection{Models and Dataset}
\label{sec:dataset}

\textbf{Models.} To ensure the comprehensiveness and depth of our experimental results, we selected a diverse set of state-of-the-art open-source and closed-source models, encompassing Video-LLMs, Audio-LLMs, LLMs, and Reasoning Models. Detailed specifications for each model, including their backbone architecture, training strategy, training data composition, open/closed-source status, and inference mechanisms, are summarized in Table~\ref{table 1}.

\textbf{Dataset.} Our study utilizes the \textbf{OV-MERD} dataset, which contains 332 multimodal samples from Chinese television dramas, movies, and interviews, with video, audio, and text modalities when available. The sample durations vary significantly, 
%BS: range --> ranging
ranging from 0.2\,s to 22.1\,s (mean 3.9\,s). Twenty samples are video-only due to their extremely short durations. These videos have an average frame rate of 24.9\,FPS. Statistical analysis reveals a rich vocabulary of 248 unique emotion terms, with each sample being annotated with an average of 3.34 labels. The labels are formatted as a list of strings, e.g., \textit{[suspicious, angry, dissatisfied, questioning]}.

% \noindent\textbf{{Annotation Characteristics.}}
% The emotion labels were generated using an LLM and subsequently verified by human annotators, ensuring high-quality, open-vocabulary annotations. 

\subsection{Evaluation Metrics}
\label{sec:evaluation_metrics}

We follow the official MER2024 evaluation protocol. To handle semantic synonymy (``angry'' and ``furious''), all predicted and ground-truth labels are grouped into semantic clusters using GPT-3.5-Turbo with a predefined prompt:

\begin{quote}
\textit{Please assume the role of an expert in the field of emotions. We provide a set of emotions. Please group the emotions, with each group containing synonyms or consistent emotional terms. Directly output the results, with each group in list format.}
\end{quote}

Let $G(\cdot)$ be the function that maps a label string $x$ to its semantic group ID. For a set of ground-truth labels $\{y_i\}_{i=1}^M$ and a set of predicted labels $\{\hat{y}_j\}_{j=1}^N$, the corresponding sets of group IDs are:
\begin{equation}
    Y = \{G(y_i) \mid y_i \in \{y_i\}_{i=1}^M\},
\end{equation}
\begin{equation}
    \hat{Y} = \{G(\hat{y}_j) \mid \hat{y}_j \in \{\hat{y}_j\}_{j=1}^N\}.
\end{equation}
We calculate set-level $\text{Precision}_\text{s}$ and $\text{Recall}_\text{s}$. $\text{Precision}_\text{s}$ indicates the number of correctly predicted labels; $\text{Recall}_\text{s}$ indicates whether the prediction covers all ground truth. The final metric $\text{F}_\text{s}$ is the harmonic mean of two metrics:
\begin{equation}
    % \text{Accuracy} = \frac{|Y \cap \hat{Y}|}{|\hat{Y}|}, \label{eq:Accuracy}
    \text{Precision}_\text{s} = \frac{|Y \cap \hat{Y}|}{|\hat{Y}|}, \ \text{Recall}_\text{s} = \frac{|Y \cap \hat{Y}|}{|Y|},
\end{equation}
\begin{equation}
    % \text{Recall} = \frac{|Y \cap \hat{Y}|}{|Y|}. \label{eq:recall}
    \text{F}_\text{s} = 2 \times \frac{\text{Precision}_\text{s} \times \text{Recall}_\text{s}}{\text{Precision}_\text{s} + \text{Recall}_\text{s}}.
\end{equation}
% \jh{pls check if this is called accuracy or precision?}

\noindent\textbf{{Implementation Details.}}
% We note that the model originally specified for evaluation, GPT-3.5-Turbo-16k-0613, was deprecated by OpenAI after September 13, 2024. Consequently, we use its official successor, GPT-3.5-Turbo, for all evaluations. To ensure measurement reliability, each experimental result was evaluated five times independently, and the final scores are the mean of these five runs.
% As GPT-3.5-Turbo-16k-0613 was deprecated after September 13, 2024, we adopted its official successor, GPT-3.5-Turbo, for all evaluations. 
To ensure stability and reliability, each experiment was repeated five times independently, the final results are reported as the mean scores.

\section{Experimental Findings}

% We begin our investigation with a thorough evaluation of MLLMs. Utilising a comprehensive testing scheme, we attempt to identify how different aspects of foundation models impact performance. We place particular emphasis on Average Score, which is critical considerations for multimodal emotion recognition targeted to real-world applications.

% In this section, we present extensive experimental results across four research directions, involving 19 LLMs and Multimodal LLMs, and provide detailed analysis and discussion.\jh{unclear what is the four directions, as for the first time mentioned, check if the below amended version works.}

The experimental results in this section are structured into four key aspects, aligning with the methodological framework established in Sections III-B through III-E. We present a comprehensive evaluation involving 19 LLMs and MLLMs, providing detailed analysis for each corresponding aspect of the benchmark.
% \zix{please double check the tense throughout this section! }

% \subsection{Best Method and Modality Fusion Strategy}
\subsection{Identifying the Optimal Model and Fusion Strategy}
\label{sec:exper_framework_modality}

% \subsubsection{Which introduced method in Figure~\ref{Figure 1} performs best in MER-OV?}
\subsubsection{Which architectural variant delivers optimal performance on MER-OV?}

\textbf{\textit{Description:}}
% As previously discussed, we introduced three methods for addressing MER-OV. 
% To evaluate the effectiveness of the three introduced methods in Figure~\ref{Figure 1}, we performed multimodal fusion of text, video, and audio, and compared performance across the three methods. \jh{repeated mentioning three methods in this sentence. Please replace with this: To benchmark the three proposed methods (Figure~\ref{Figure 1}), we evaluated their performance in a consistent trimodal fusion setting integrating text, video, and audio.}
To benchmark the three proposed methods (Figure~\ref{Figure 1}), we evaluated their performance in a consistent trimodal fusion setting integrating text, video, and audio.
To ensure the generalizability of our results, comparative experiments were conducted using various LLMs and Video-LLMs.

% \textbf{\textit{Answer: }}\textbf{Emotional clue-based two-stage method achieves the highest performance, follows\jh{followed} by Video-LLM-only-based one-stage\jh{ method}, while objective description-based two-stage method exhibits relatively poorer results.}
\textbf{\textit{Answer: }}\textbf{The emotional clue-based two-stage method achieves the highest performance, followed by the Video-LLM-only-based one-stage method, while the objective description-based two-stage method exhibits relatively poorer results.}

\begin{figure}[h]
  \centering
  \includegraphics[width=0.5\textwidth]{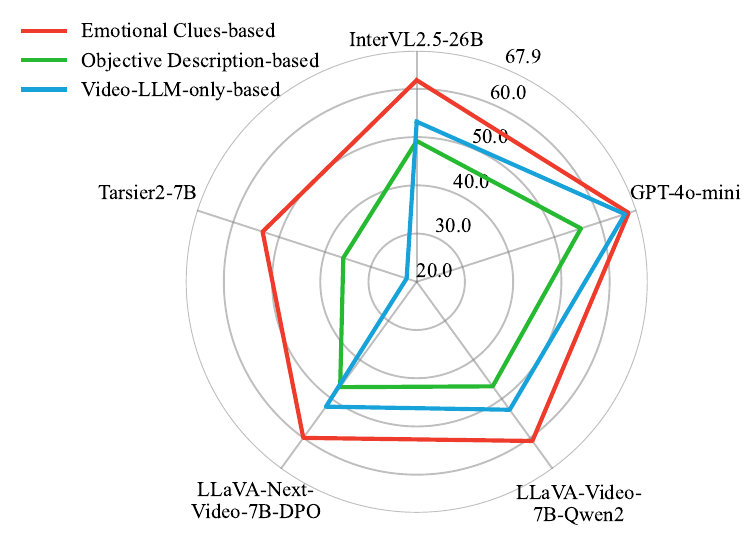}
  % \caption{Performance comparison of three Video-LLMs under multimodal clues, multimodal description, and Video-LLM prediction settings. The Audio-LLM used is Qwen2-Audio-7B, and the LLM is Gemma2-9B.} 
  % \caption{Performance comparison among \textit{three LLM-based} approaches, \ie emotional clue-based two-stage methods (red), objective description-based two-stage methods (blue), and Video-LLM-only-based one-stage methods (green), considering the \textit{open-source} Audio/Video/-LLMs only. In the first stage, one Audio-LLM, \ie Qwen2-Audio-7B, and four different Video-LLMs, \ie LLaVA-video-7B-Qwen2, Tarsier-7B, LLaVA-NeXT-Video-7B-DPO, and InternVL2.5-26B are employed. In the second stage, four LLMs, \ie Gemma2-9B, Llama3.1-8B, Qwen2.5-7B, and Qwen2.5-32B, are used for evaluation to reduce the performance fluctuation.} 
    \caption{Performance comparison among \textit{three LLM-based} approaches, \ie emotional clue-based two-stage method (red), objective description-based two-stage method (blue), and Video-LLM-only-based one-stage method (green), evaluated using  $\text{F}_\text{s}$ score.
    % , considering the \textit{open-source} Audio/Video/-LLMs only. 
    In the first stage, one Audio-LLM, \ie Qwen2-Audio-7B, and five different Video-LLMs in Table~\ref{table 1} are employed. In the second stage, four LLMs, \ie Gemma2-9B, Llama3.1-8B, Qwen2.5-7B, and Qwen2.5-32B, are used for evaluation to reduce the performance fluctuation.}
    % \jh{what is the metrics here? F1} }
  % \zix{1. Add `-based' to the first and second titles in the legends. 2. Better not to use bolden font if they are not emphasised. Same with the other figures.}}
  \label{Figure 3}
\end{figure}

\textbf{\textit{Details: }}Figure \ref{Figure 3} presents a comparative analysis of the three methods. Both, the emotional clue-based two-stage method and objective description-based two-stage method integrate trimodal processing, employing five Video-LLMs for video analysis, Qwen2-Audio-7B for audio processing, and four open-source LLMs for text comprehension.
% With a fixed Audio-LLM, we systematically evaluated all 20 possible combinations of the five Video-LLMs and four LLMs, subsequently averaging results per Video-LLM across LLM variations as baseline comparisons.

This superiority arises from the mechanism of the emotional clue-based two-stage method, wherein Video/Audio-LLMs extract emotionally salient clues via their reasoning capabilities, which are subsequently synthesized by the text-based LLMs. In contrast, the objective description-based two-stage method relies on objective multimodal descriptions, which, although containing primary emotional clues, fail to adequately capture subtle affective states due to existing limitations of LLMs. This manifests in comparable $\text{Precision}_\text{s}$ scores but significant disparities in $\text{Recall}_\text{s}$, ultimately leading to substantial differences in overall performance.

The Video-LLM-only-based one-stage method exhibits the opposite pattern, with higher $\text{Precision}_\text{s}$ but lower $\text{Recall}_\text{s}$, indicating substantial omissions of emotional states. These findings suggest that although contemporary open-source Video-LLMs exhibit competent video comprehension abilities, they remain inferior to general-purpose LLMs for complex multimodal reasoning tasks involving nuanced affective understanding.

\subsubsection{What is the optimal modality fusion strategy?}

% \begin{flushleft}

\textbf{\textit{Description: }}
% Prior research and empirical evidence indicate that different modalities provide complementary information, and their integration generally enhances model performance. 
To evaluate the benefits of multimodal fusion, we conducted comparative experiments with the following configurations: (1) text and video fusion, (2) text and audio fusion, and (3) text, video, and audio fusion.

\textbf{\textit{Answer: }}\textbf{The trimodal configuration outperforms all bimodal and unimodal setups. Furthermore, the video modality demonstrates a more significant impact on performance than both the audio and text modalities.}

\textbf{\textit{Details: }}In our experiments, Qwen2-Audio-7B was used as the audio processing LLM, and Gemma2-9B served as the general-purpose LLM. For visual understanding, we selected five Video-LLMs in Table~\ref{table 1}. This selection enabled comprehensive comparison between architectural approaches and development paradigms, while maintaining consistent audio-textual processing across all experimental conditions.

As shown in Table \ref{table 2}, unimodal performance varies considerably across modalities. Both text (54.8\%) and video (55.6\%) achieve relatively strong results, while audio (47.2\%) lags behind, highlighting that linguistic and visual cues are more discriminative for emotion recognition in MER-OV.
Bimodal fusion consistently improves over unimodal settings. In particular, text + video outperforms either modality alone, confirming that these two carry complementary information. Although audio performs poorly in isolation, its integration with text or video still leads to measurable gains, suggesting that acoustic signals provide useful auxiliary cues rather than serving as a dominant modality.
The trimodal fusion further advances performance, achieving 61.0\%, which surpasses the best bimodal result by a clear margin. Beyond numerical improvement, this setting demonstrates enhanced robustness across $\text{Precision}_\text{s}$ and $\text{Recall}_\text{s}$, showing that leveraging all three modalities enables a more comprehensive representation.% of emotional states.

\begin{table}[t]
\centering
% \caption{Using different modal clues to predict emotional states and comparing the impact of different modal clues on performance.}
\caption{Evaluation of the \textit{modality contribution} by various combinations of the text, audio, and video modalities with the emotional clue-based methods. Specifically, only one LLM -- Gemma2-9B and one Audio-LLM -- Qwen2-Audio-7B, but five Video-LLMs are considered, as the video modality plays a dominant role
% \jh{it is not clear when 5 vllm was evaluated, how is the performance here show this, it is averaged across 5 vllms?}, \textcolor{blue}{gzq: add content: and the average result of the five Video-LLMs is reported.}
}
\label{table 2}
\begin{tabular}{cccccc}
\toprule
Text & Video & Audio & $\text{Precision}_\text{s}$ [\%] & $\text{Recall}_\text{s}$ [\%] & $\text{F}_\text{s}$ [\%] \\
\midrule
% Gemma2-9B & $\times$ & $\times$ & 57.8 & 52.3 & 55.0 \\
% Gemma2-9B & $\times$ & Qwen2-Audio-7B & 58.4  & 47.3 & 52.8 \\
$\checkmark$ & $\times$     & $\times$     & 57.8 & 52.3 & 55.0 \\
$\times$     & $\checkmark$ & $\times$     & \textbf{59.5} & \textbf{55.8} & \textbf{57.6} \\
$\times$     & $\times$     & $\checkmark$ & 48.1 & 46.3 & 47.2 \\
\midrule
$\checkmark$ & $\checkmark$ & $\times$     & \textbf{60.2} & \textbf{55.0} & \textbf{57.5}\\
$\checkmark$ & $\times$     & $\checkmark$ & 58.4 & 52.8 & 55.5 \\
$\times$     & $\checkmark$ & $\checkmark$ & 58.8 & 53.6 & 56.1\\
\midrule
$\checkmark$ & $\checkmark$ & $\checkmark$ & \textbf{60.7} & \textbf{61.4} & \textbf{61.0}\\

\bottomrule
\end{tabular}
\end{table}

% \subsection{Best Performance of Video-LLMs, Audio-LLMs, and LLMs\jh{How about: A Deeper Look into Modality-wise Model Performance}}
\subsection{A Deeper Look into Modality-wise Model Performance}
\label{sec:exper_model_benchmark}

% In summary, the results indicate that text and video are the most informative modalities, while audio contributes supportive but valuable information. Importantly, full multimodal fusion proves to be the most effective strategy, underscoring the necessity of exploiting cross-modal complementarities in MER-OV.\jh{A bit strange to see a first very beginning paragraph of a section starting with In summary... Pls check if the following works better, the logic flow is starting from the findings on all three, and then move to the seperate, which is quite different from the current one which start from each single one and then combine: Our findings demonstrate that a complete trimodal fusion strategy yields optimal performance. Moreover, a modality-wise decomposition indicates that text and video are the primary drivers of this performance, with audio delivering supplementary gains.}
Our findings demonstrate that a complete trimodal fusion strategy yields optimal performance. Moreover, a modality-wise decomposition indicates that text and video are the primary drivers of this performance, with audio delivering supplementary gains.
\subsubsection{What is the optimal Video-LLM in Stage 1?}

% \begin{flushleft}

\textbf{\textit{Description: }}
We conducted experiments within the emotional clue-based two-stage method to compare the performance of five Video-LLMs for the MER-OV task. The Audio-LLM was held constant (Qwen2-Audio-7B) while we tested five different Video-LLMs. Each Video-LLM was paired with seven different LLMs, and its final performance was presented as the mean score across these seven runs.

\textbf{\textit{Answer: }}\textbf{Closed-source GPT-4o-mini outperforms other larger open-source models; preference alignment techniques are an effective training strategy for Video-LLMs.}
% \zix{it is unclear of the second half of this sentence}.

\begin{figure}[t]
  \centering
  \includegraphics[width=0.5\textwidth]{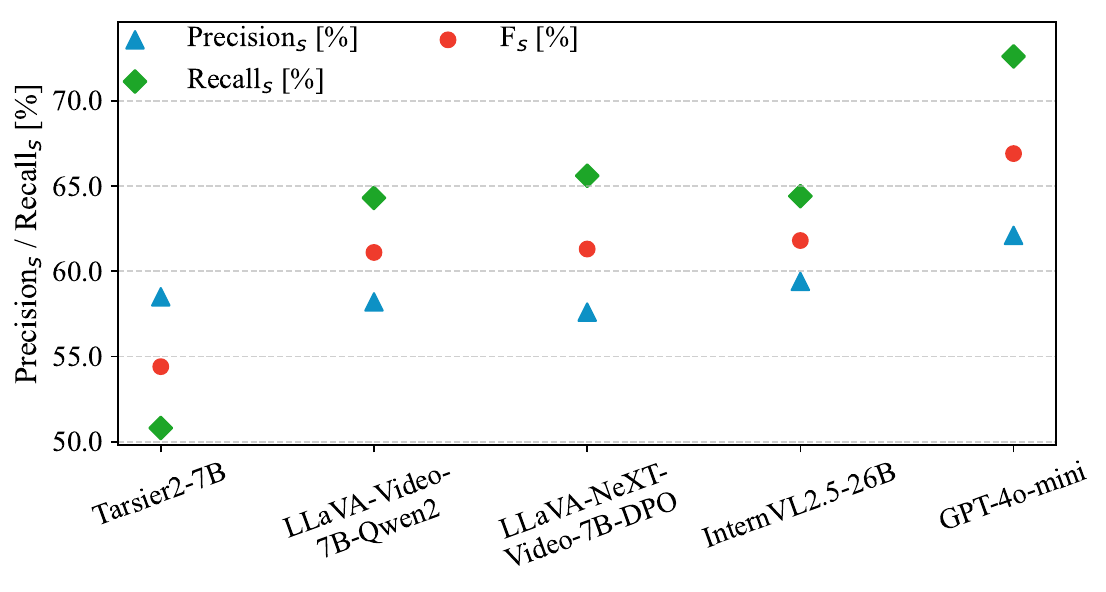} 
  % \caption{Performance comparison among \textit{five open/closed source Video-LLMs (four on-premises and one API request)} with emotional clue-based methods. The Audio-LLM used is Qwen2-Audio-7B, while the \textit{eight open/closed source LLMs (four on-premises and four API request)} are included in the comparison. The horizontal axis represents accuracy, while the vertical axis represents recall, these values are mean results of \textit{the eight open/closed-source different LLMs} are reported.}
  \caption{Performance comparison among \textit{five open/closed-source Video-LLMs} evaluated with emotional clue-based methods. Audio-LLM: Qwen2-Audio-7B; LLMs: seven open/closed-source LLMs, whose mean scores form the performance points.}
  \label{Figure 5}
\end{figure}

\textbf{\textit{Details: }}As shown in Figure~\ref{Figure 5}, among the open-source Video-LLMs, InternVL2.5-26B, LLaVA-NeXT-Video-7B-DPO, and LLaVA-Video-7B-Qwen2 exhibit comparable performance, while Tarsier2-7B lags behind. The closed-source GPT-4o-mini outperforms these larger open-source models, indicating that proprietary optimizations enable superior handling of temporal dynamics and fine-grained visual cues in MER-OV without relying on massive parameter counts. This challenges the parameter-performance proportionality in multimodal models, underscoring the primacy of architectural efficiency and training data alignment over scale.

Furthermore, the performance parity among top Video-LLMs—spanning a 3.7× parameter difference—reinforces that size alone is insufficient. The superiority of LLaVA-NeXT-7B-DPO over LLaVA-7B-Qwen2, sharing identical language backbones, demonstrates the critical role of alignment techniques like Direct Preference Optimization (DPO)~\cite{DPO} in elevating base model performance, particularly for temporal reasoning in emotion clue extraction. Recent advancements, such as Modified Preference Optimization (MPO) in InternVL2.5 to mitigate spatial hallucinations and hybrid SFT+RLHF pipelines, further validate DPO's efficacy in state-of-the-art implementations for LLaVA-family models.

% \subsubsection{How do diverse Audio-LLMs perform in Stage 1 of MER-OV?\jh{same as above: What is the optimal Audio-LLM in Stage 1? }}
\subsubsection{What is the optimal Audio-LLM in Stage 1? }

\textbf{\textit{Description: }}We conducted experiments within the emotional clue-based two-stage method to compare the performance of five Audio-LLMs for the MER-OV task. The Video-LLM was held constant (GPT-4o-mini) while we tested five different Audio-LLMs. Each Audio-LLM was paired with seven different LLMs, and its final performance was presented as the mean score across these seven runs.

\textbf{\textit{Answer: }}\textbf{Closed-source Gemini's progressive versions show gains from enhanced alignment and acoustic training for subtle clues. Open-source Qwen2-Audio-7B matches Gemini-2-pro.}

\textbf{\textit{Details: }}We evaluated five Audio-LLMs from two series: Qwen-Audio and Gemini. All models are listed in Table~\ref{table 1}.

\begin{figure}[t]
  \centering
  \includegraphics[width=0.5\textwidth]{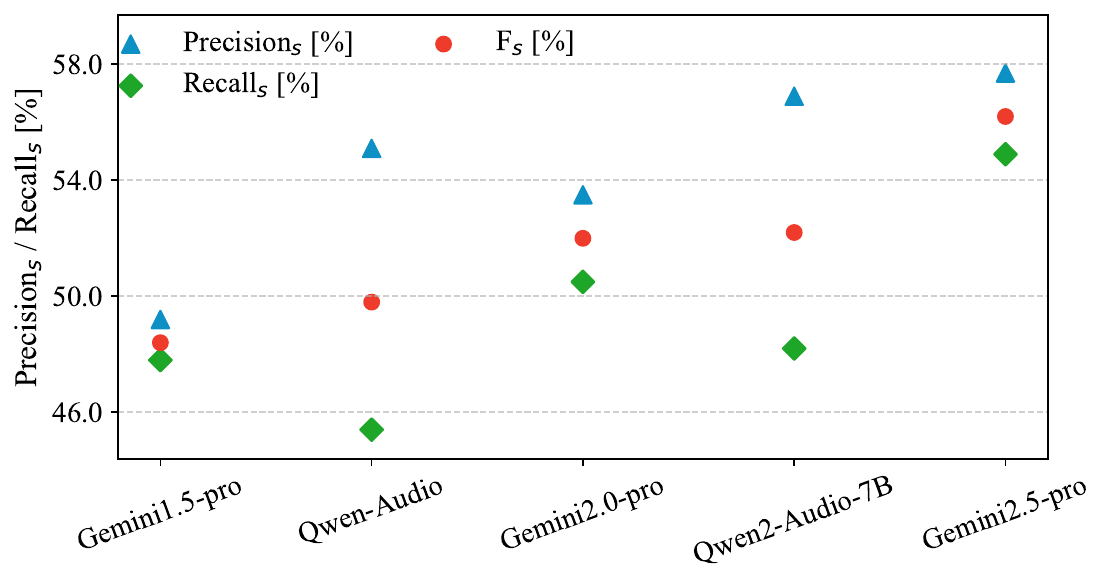} 
  \caption{Performance comparison among \textit{five open/closed-source Audio-LLMs} with emotional clue-based method. LLMs: seven open/closed-sourced LLMs, whose mean scores form the final performance. The Video-LLM (GPT-4o-mini) is held constant to reduce its influence on the final performance.
  }
  \label{Figure 6}
\end{figure}

As shown in Figure \ref{Figure 6}, Gemini2.5—the most advanced MLLM from Google at the time of writing—demonstrates superior audio analysis capabilities compared to the open-source Qwen-Audio models. It consistently outperforms other models in both $\text{Precision}_\text{s}$ and $\text{Recall}_\text{s}$. In the audio modality, emotion clues primarily derive from tone, pitch, speed, and pauses. However, compared to video, the emotional information captured from audio is quite limited. This limitation is reflected in the generally shorter emotion word lists generated using audio clues compared to those derived from video clues. Furthermore, the audio modality presents certain limitations in our dataset: among the 332 MER-OV samples, 20 samples (6\%) lack audio data, which constrains the upper bound of audio+text-based emotion recognition system performance.
% \jh{??? hard to follow why it constrains the upper bound?}\textcolor{blue}{gzq: Due to the lack of audio information for these 20 pieces of data, their corresponding text content is also blank. This makes it difficult for the "text + audio" recognition scheme to accurately identify this part of the data, which in turn leads to extremely low prediction accuracy of these 20 pieces of data and directly drags down the overall indicators. It should be clarified that this does not limit the upper limit; instead, it inherently results in a relatively low indicator for the text + audio scheme.
Overall, the current availability of high-quality Audio-LLMs remains limited, and these results highlight a clear opportunity for further improvement in this area.

% \subsubsection{How do diverse LLMs perform in Stage 2 of MER-OV?\jh{What is the optimal LLM in Stage 2?}}
\subsubsection{What is the optimal LLM in Stage 2?}
% \begin{}

\textbf{\textit{Description: }}Although many benchmarks exist for evaluating LLMs, systematic comparisons on MER-OV remain scarce. To fill this gap, we evaluate seven LLMs (cf. Table \ref{table 1}) using an emotional clue-based two-stage method. The Audio-LLM (Qwen2-Audio-7B) was fixed, while seven different LLMs were tested, each paired with five Video-LLMs. The final performance was reported as the mean score across these runs.

% Although numerous benchmarks have been proposed to evaluate LLM capabilities, systematic comparisons specifically targeting their performance on MER-OV tasks remain limited. To address this gap, we selected seven open-source and closed-source LLMs as shown in Table \ref{table 1} by using emotional clue-based two-stage method. The Audio-LLM was held constant (Qwen2-Audio-7B) while we tested five different LLMs. Each LLM was paired with five different Video-LLMs, and its final performance is presented as the mean score across these five runs.

% \begingroup
% \tolerance=2000
% \emergencystretch=3em
% \textbf{\textit{Answer: }}\textbf{For open-source text LLMs, the performance ranking is as follows: Llama3.1-8B is ranked higher than DeepSeek-V3, which in turn outperforms Qwen2-72B. Following Qwen2-72B is Gemma2-9B, then Qwen2.5-32B, and finally Qwen2.5-7B.}
\textbf{\textit{Answer: }}\textbf{In MER-OV, where nuanced emotional inference is key, optimized architectures and pre-training for semantic richness outweigh model size. Variability within families (e.\,g., Qwen series) indicates that iterative refinements enhance affective generation.}

\begin{figure}[t]
  \centering
  \includegraphics[trim={0cm 0cm 0cm 0cm}, clip, width=0.5\textwidth]{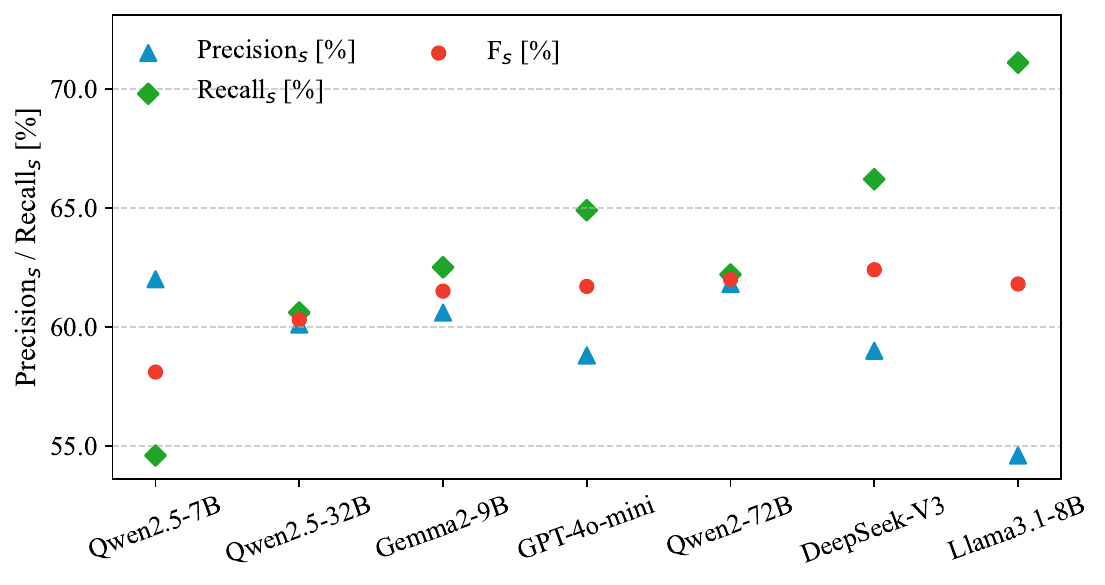}
  % \caption{Performance comparison of eight LLMs using multimodal clue methods. The Audio-LLM used is Qwen2-Audio-7B, while the five evaluated Video-LLMs are included in the comparison. The vertical axis represents the mean value of these five Video-LLMs. }
  % \caption{Performance comparison among \textit{eight open/closed-source different LLMs (four on-premises and four API request)} with emotional clue-based methods. The Audio-LLM used is Qwen2-Audio-7B, while the \textit{five evaluated open/closed-source Video-LLMs (four on-premises and one API request)} are included in the comparison. The vertical axis represents the mean value of these five Video-LLMs.}
  \caption{Performance comparison among \textit{seven open/closed-source LLMs}  evaluated with emotional clue–based methods. Audio-LLM: Qwen2-Audio-7B; Video-LLMs: five open/closed-source models, whose mean scores form the vertical axis.}
  \label{Figure 4}
\end{figure}

\textbf{\textit{Details: }}
% As shown in Figure \ref{Figure 4}, the open-source Llama3.1-8B surprisingly outperformed all Qwen variants, despite having a significantly smaller parameter size (8B vs. 72B or DeepSeek-V3). Larger-scale models, such as DeepSeek-V3 and Qwen2-72B, maintained competitive advantages over other open-source alternatives. The closed-source models GPT-4o-mini and Gemini2.0-flash achieved performance comparable to Qwen2-72B. Notably, the performance gap between open-source and closed-source models remains marginal. The suboptimal performance of Qwen2.5-7B and Qwen2.5-32B may be due to their predominantly Chinese training corpus, which could limit their effectiveness on English-oriented tasks compared to models primarily trained on English data. Llama3.1-8B achieved the highest recall among all models. Although it had the lowest accuracy, it still obtained the highest average score. Analysis of its outputs revealed that Llama3.1-8B generates long emotion lists, averaging eight emotion words, with some redundancy among the terms. Gemini2.0-flash exhibits a similar tendency.
As shown in Figure \ref{Figure 4}, the open-source Llama3.1-8B outperforms all Qwen variants despite its smaller size (8B vs. 72B/DeepSeek-V3). Larger models such as DeepSeek-V3 and Qwen2-72B remain competitive, while closed-source GPT-4o-mini performs on par with Qwen2-72B, showing only a marginal gap between open- and closed-source models. In contrast, Qwen2.5-7B and Qwen2.5-32B underperform, likely due to their Chinese-focused training data. Llama3.1-8B achieves the highest $\text{Recall}_\text{s}$ and overall $\text{F}_\text{s}$, though its $\text{Precision}_\text{s}$ is lower. Further analysis suggests that it is due to its tendency to generate longer, redundant emotion lists.

Gemma2-9B, Qwen2.5-32B, and Qwen2-72B show balanced $\text{Precision}_\text{s}$ and $\text{Recall}_\text{s}$ (difference less than 2.0) and generate concise sentiment lists of three to four words, indicating precise predictions. In contrast, Qwen2.5-7B Instruction is the only model with higher $\text{Precision}_\text{s}$ than $\text{Recall}_\text{s}$; its shorter outputs are accurate but less comprehensive for open-vocabulary tasks. Most other models display higher $\text{Recall}_\text{s}$ than $\text{Precision}_\text{s}$, typically between 60.0\% and 70.0\%. %BS: added \% after 60.0 and 70.0 
For MER-OV, it is important to balance $\text{Precision}_\text{s}$ with a comprehensive description of the psychological state, ensuring that each word in the emotion list is precise and non-redundant.
% The accuracy and recall of Gemma2-9B, Qwen2.5-32B, and Qwen2-72B are more balanced, with differences not exceeding 0.020. Their sentiment list outputs range from three to four words, indicating relatively precise sentiment predictions. However, Qwen2.5-7B Instruction is unique, being the only LLM with accuracy exceeding recall. Its sentiment lists are the shortest; while the predictions are more accurate, they lack the comprehensiveness required for open vocabulary tasks. Other models generally exhibit higher recall than accuracy, typically ranging from 0.600 to 0.700.

% For the MER-OV task, it is desirable to ensure high accuracy while providing a sufficiently comprehensive description of the psychological state. Ideally, each word in the emotion list should accurately describe the current emotional state without redundancy.

\begin{figure*}[h]
  \centering
  \includegraphics[width=1\textwidth]{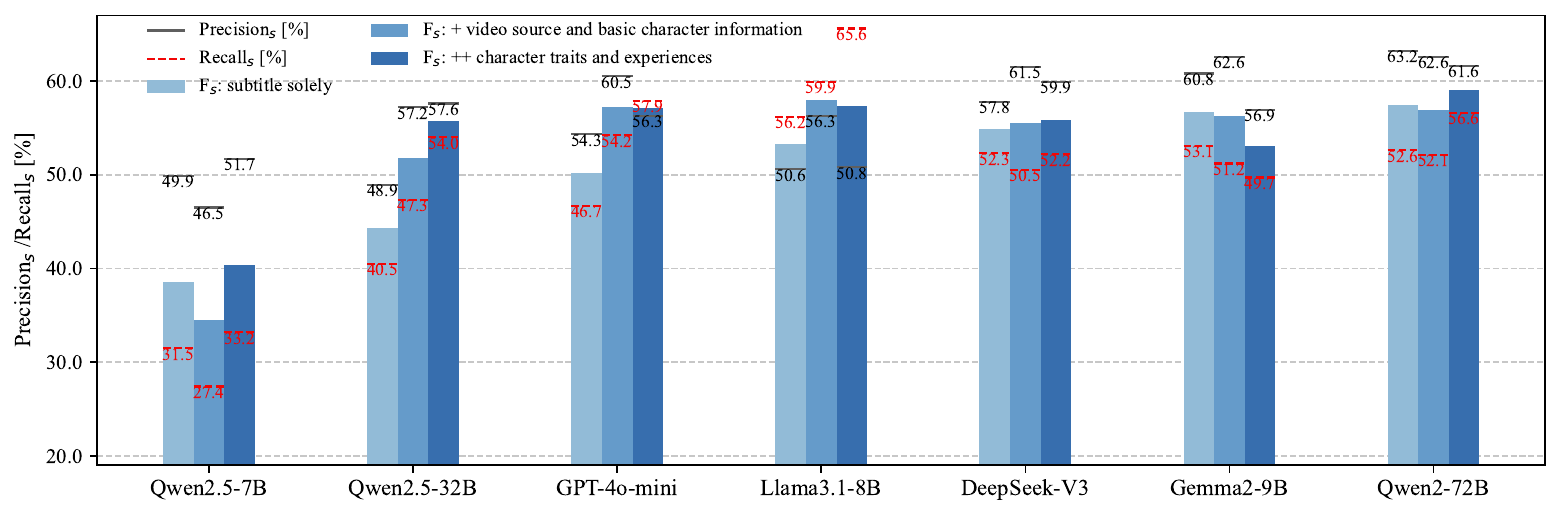} 
  % \caption{Performance comparison based on the text modality with contextual information. Different content in prompt: (1) inferring emotional states based solely on subtitle content; (2) inferring emotional states with the addition of video source and character information; and (3) inferring emotional states by further incorporating character traits and experiences alongside video source and character information. We used five different LLMs to evaluate.}
  \caption{Evaluation of \textit{additional textual context information} on the performance of the \textit{LLM-only-based one-stage method} with all seven selected LLMs. Different prompting strategies are used: (1) subtitle only; (2) subtitle + video source and basic character information; and (3) subtitle + video source and basic character information + character traits and experiences.
  % \jh{from the caption and the figure, I cannot see which results are recall? But in text you said recall shows some improvement?}.\textcolor{blue}{gzq: This figure has been updated}
  } %\textcolor{cyan}{Method: LLM-only-based one-stage method (only based subtitle and additional information, utilize LLM to predict emotional states). Audio-LLM: no used; Video-LLM: no used; LLMs:  eight open/closed-source LLMs (four on-premises and four API requests). We will add the results of three LLMs.}}}
  \label{Figure 8}
\end{figure*}

\subsection{Video Modality Processing Techniques}
\label{sec:exper_video_enhancement}

% \subsubsection{Can a performance enhancement be achieved by adding video metadata?\jh{What is the role of video metadata in performance enhancement?}}
\subsubsection{What is the role of video metadata in performance enhancement?}

% \begin{flushleft}
\textbf{\textit{Description: }}Emotional states depend on both events and cognition, with the latter shaped by personality and past experiences. To capture this, prompts were enriched with contextual details such as video content and character information (e.\,g., names, traits, background). To avoid confounding factors, this experiment relied only on the text modality (subtitles).
% Emotional states are influenced by both events and cognitive factors. Events refer to occurrences within the video, while cognition develops over time and is closely related to a character’s personality and past experiences. Therefore, enriching prompts with additional contextual information—such as video content and character-related details (e.g., names, personality traits, and personal background)—may help LLMs better understand the cognitive states of characters depicted in the video. To minimize confounding variables, this experiment used only the text modality (subtitle content).
% Seven LLMs was selected as shown in Table \ref{table 1}.

\textbf{\textit{Answer: }\textbf{Introducing video sources and character names improves overall $\text{Precision}_\text{s}$ and $\text{Recall}_\text{s}$. However, further adding character past experiences and personality traits leads to a decrease in $\text{Precision}_\text{s}$, while $\text{Recall}_\text{s}$ shows some improvement.
% jh{did not see results in terms of recall?}\textcolor{blue}{gzq: Figure 8 has been added recall.}
}}

\textbf{\textit{Details: }}As shown in Figure \ref{Figure 8}, after introducing video sources and character information, $\text{Precision}_\text{s}$ improves in five of the seven LLMs,
% \jh{why three out of four? I thought you have seven LLMs in total? also in Fig 8 there are seven in total?}\textcolor{blue}{gzq: It should change: As shown in Figure \ref{Figure 8}, after introducing video sources and character information, $\text{Precision}_\text{s}$ improves in five of the seven LLMs.}
while Qwen2-72B shows a slight decrease. The improvement is most pronounced in models with lower initial performance.These results indicate that LLMs can effectively leverage video sources and character names for inference, with $\text{Recall}_\text{s}$
% \jh{which is recall?} \textcolor{blue}{gzq: Figure 8 has been added recall.} 
also showing corresponding improvements.

When character past experiences and personality traits were further introduced, three LLMs exhibit decreased $\text{Precision}_\text{s}$ but slight improvements in $\text{Recall}_\text{s}$.
% \jh{which is recall in fig 8}\textcolor{blue}{gzq: Figure 8 has been added recall.}. 
This may be because character experiences and personality traits often contain expressions related to emotional states, which may draw the LLM’s attention and interfere with accurate emotional state inference.

\subsubsection{To what extent do frame sampling strategies impact MER-OV performance?}

% \begin{flushleft} 

\textbf{\textit{Description: }}In earlier experiments, video clues were generated by uniformly sampling 24 frames per video. Given the varying video lengths in the dataset, we adopted a dynamic frame sampling strategy with rates of 1, 2, 4, and 6 fps. The Audio-LLM (Qwen2-Audio-7B) and LLM (Llama3.1-8B) were fixed, while five Video-LLMs (Table \ref{table 1}) were evaluated.

\textbf{\textit{Answer: }}\textbf{Compared to fixed-frame sampling, dynamic frame sampling consistently yields better performance. The optimal sampling rate, however, varies among different Video-LLMs.}

\begin{figure}[h]
  \centering
  \includegraphics[width=0.51\textwidth]{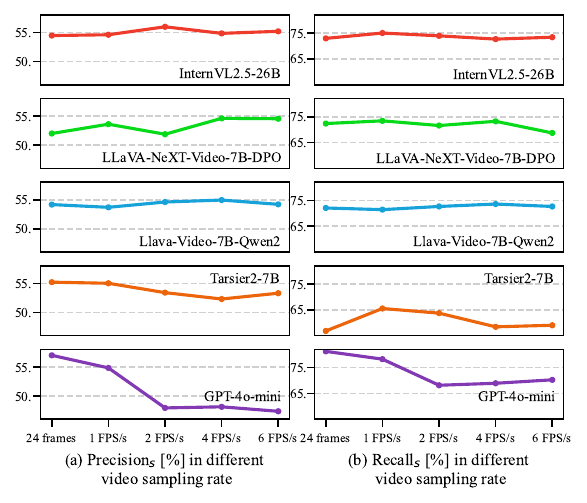} 
  % \caption{Performance Comparison of the different video frame extraction strategy. The selected Video-LLMs are InternVL2.5-26B, LLaVA-NeXT-Video-7B-DPO, and LLaVA-Video-7B-Qwen2. The strategies are uniformly sampling 24 frames from each video and dynamic sampling frame per second(1 frame/s, 2 frame/s, 4 frame/s, and 6 frame/s).}
  \caption{Performance evaluation of \textit{video sampling rate for the Video-LLMs} with the emotional clue-based methods, where five open-source Video-LLMs are employed. Two uniformly sampling strategies are considered: fixed sampling frames over each video (\ie 24 frames) and dynamic sampling rates (\ie 1, 2, 4, or 6 frames/s).} 
  % \textcolor{green}{Audio-LLM: Qwen2-Audio-7B; LLM: Llama3.1-8B.}} 
  \label{Figure 7}
\end{figure}

\textbf{\textit{Details: }}Figure \ref{Figure 7} shows the performance of the three models under each sampling method. InternVL2.5-26B achieved optimal performance at 2 FPS: while $\text{Recall}_\text{s}$ slightly decreases compared to 1 FPS, $\text{Precision}_\text{s}$ improves %BS: Do not use "significantly" w/o naming test method and p-value - I changed for:
considerably. 
Performance decline when the frame rate exceeded this value. Both LLaVA-NeXT-Video-7B-DPO and LLaVA-Video-7B-Qwen2 achieve their highest $\text{Recall}_\text{s}$ and $\text{Precision}_\text{s}$ at 4 FPS; further increases in sampling rate results in performance degradation. Notably, LLaVA-NeXT-Video-7B-DPO also performs well at 1 FPS, whereas LLaVA-Video-7B-Qwen2 performed poorly under the same condition.

Overall, dynamic frame sampling based on video duration outperforms fixed-frame sampling. For shorter videos, fixed-frame sampling may introduce redundancy, distracting the Video-LLM and impairing its ability to extract meaningful features. As Video-LLMs do not prescribe a specific number of input frames, determining the frame count requires empirical evaluation for each model to achieve optimal performance.
% \end{flushleft}

% \section{Prompt Engineering}
\subsection{Prompt Engineering}
\label{sec:exper_prompting}
In this section, we systematically investigate the impact of different prompting strategies on the performance of LLMs in the Open-Vocabulary Emotion Inferencing stage. 
% Our experimental design adopts a hierarchical approach, progressing from basic to advanced prompting techniques. 
% Specifically, we divide our investigation into three stages. 
% First, we analyze \textbf{hard prompt design patterns}, detailing the construction of input prompt templates for the Open-Vocabulary Emotion Inferencing stage. Second, we examine \textbf{composite prompting strategies}, which extend the previous designs by integrating various optimization techniques to enhance model performance.
% Third, we explore \textbf{meta-prompting}, a key APE approach distinguished by its task-agnostic and automated nature. 
% Lastly, we evaluate \textbf{Reasoning Models} that incorporate advanced reasoning mechanisms, aiming to push the boundary of emotion inference in open-vocabulary settings.

% \subsubsection{Impact of Hard Prompt Design Patterns on Model Performance in MER-OV task.}
% \subsubsection{What impact do hard prompt design patterns have on model performance in MER-OV?\jh{How do hard prompt design choices influence MER-OV performance?}}
\subsubsection{How do hard prompt design choices influence MER-OV performance?}
% \begin{flushleft}

\textbf{\textit{Description: }}
% The formulation of prompts plays a crucial role in unlocking the capabilities of LLMs. 
We investigated the impact of different hard prompt design patterns on the performance of various LLMs in MER-OV. The corresponding prompt content is integrated with the multimodal information obtained in \textit{Stage 1} to form the input instructions for \textit{Stage 2}, as illustrated in Figure \ref{overview}. 
% Five prompting strategies are compared:

% \begin{itemize}

%     \item \textbf{Standard Baseline (STD)}: A prompt that only contains a basic task description, similar to a problem statement (without any guiding or prompting information).
%     \item \textbf{Zero-shot Chain-of-Thought (Zero-shot-CoT)}: An extension of STD by appending the step-by-step reasoning trigger ``Let's think step by step.'' to guide structured problem-solving.
%     \item \textbf{Handcrafted Zero-shot}: A manually designed prompt that incorporates domain-specific heuristics, utilizing common prompt-writing techniques such as specifying roles, guiding attention to key points, and providing simple processing methods.
%     \item \textbf{Handcrafted Few-shot}: A one-shot prompt that uses synthetic examples (due to the lack of in-task training data) to simulate plausible input-output pairs, combined with handcrafted instructions.
%     \item \textbf{Multipersona}: Directs the LLM to simulate multiple ``experts'', analyzing the problem from distinct expert perspectives and offering suggestions, ultimately synthesizing a final answer. This process is completed in a single round of interaction.

% \end{itemize}

% As shown in Appendix \ref{appendix:prompts}, the specific prompt content is provided.

% \textbf{\textit{Answer: }}\textbf{The different prompting methods exhibit varying degrees of adaptability across different LLMs, but generally follow the pattern: STD \textless\ Zero-shot-CoT \textless\ Handcrafted Zero-shot \textless\ Handcrafted Few-shot.} 
\textbf{\textit{Answer: }}\textbf{For complex tasks, basic prompts offer minimal gains with superficial outputs, while customized handcrafted methods boost adaptability, align nuanced descriptions, and mitigate biases.}

% Moreover, the performance of the Multipersona method shows considerable variation across different LLMs, and may be better suited for LLMs with stronger overall capabilities.

\textbf{\textit{Details: }}The experimental results shown in Table~\ref{table 3} indicate that different combinations of LLMs and hard prompting strategies yield varied results, with each model showing a preference for different designs, making it difficult to observe a clear pattern. However, from the perspective of hard prompt design patterns, the general trend aligns with the order: STD \textless\ Zero-shot-CoT \textless\ Handcrafted Zero-shot \textless\ Handcrafted Few-shot, with the Multipersona method excluded from the overall pattern due to significant performance variations across different models. Notably, Llama3.1-8B performs excellently in the first three methods, largely due to its high $\text{Recall}_\text{s}$, meaning it predicts a greater number of emotional state words on average. In contrast, the performance of the last two methods decreases, possibly due to the limited number of example labels and the more complex analysis process. Furthermore, larger parameter models, whether open-source or closed-source, generally outperform smaller parameter open-source models.

% \end{flushleft}

\begin{table}[h]  
\centering
\setlength{\tabcolsep}{4pt} % 默认值是 6pt
% \caption{The performance of different hard prompt design patterns on different LLMs.The underlined portions in the table represent the best performance within each group, while the bolded portions indicate the best overall performance across all groups.}
\caption{Performance comparison among \textit{different hard prompt designs} for seven selected LLMs with the emotional clue-based methods. Audio-LLM: Qwen2-Audio-7B; Video-LLM: GPT-4o-mini. The underlined and bolded values indicate the best results within each group or across all groups.
% \jh{pls check why in handcrafted few-shot, there are 3 underlined values, will it be better to also highlight best results in terms of accuracy and recall as well?}\textcolor{blue}{gzq: highlighted best result in precision and recall}
} %\textcolor{cyan}{Method: emotional clue-based method. Audio-LLM: Qwen2-Audio-7B; Video-LLM: GPT-4o-mini; LLMs: six open/closed-source LLMs (three on-premises and three API requests). In the second stage, five different hard prompt designs.}}
\label{table 3}
\begin{tabular}{c|c|ccc}
\toprule
Hard Prompt & LLM & $\text{Precision}_\text{s}$ [\%]  & $\text{Recall}_\text{s}$ [\%] & $\text{F}_\text{s}$ [\%] \\
\midrule
\multirow{5}{*}{STD} 
    & Gemma2-9B & \underline{63.5} & 69.4 & 66.3 \\
    & Llama3.1-8B & 60.6 & 76.1 & 67.4 \\
    & Qwen2.5-7B & 62.1 & 69.6 & 65.6 \\
    & Qwen2.5-32B & 61.6 & \underline{76.6} & \underline{68.3} \\
    & Qwen2-72B & 60.7 & 76.1 & 67.5 \\
    & DeepSeek-V3 & 59.9 & 74.6 & 66.4 \\
    & GPT-4o-mini & 58.8 & 75.0 & 65.9 \\
\midrule
\multirow{5}{*}{Zero-shot-CoT} 
    & Gemma2-9B & \underline{64.5}  & 68.7 & 66.5 \\
    & Llama3.1-8B & 61.6 & \underline{75.9} & \underline{68.0} \\
    & Qwen2.5-7B & 63.5 & 69.4 & 66.3 \\
    & Qwen2.5-32B & 59.4 & 74.4 & 66.1 \\
    & Qwen2-72B & 60.7 & 75.3 & 67.2 \\
    & DeepSeek-V3 & 60.6 & 75.6 & 67.3 \\
    & GPT-4o-mini & 60.6 & 74.2 & 66.7 \\
 \midrule
\multirow{5}{*}{\shortstack{Handcrafted\\\\Zero-shot}} 
    & Gemma2-9B & 63.3 & 69.1 & 66.1 \\
    & Llama3.1-8B & 57.0 & \textbf{\underline{81.3}} & 67.1 \\
    & Qwen2.5-7B & \underline{65.8} & 60.7 & 63.1 \\
    & Qwen2.5-32B & 62.9  & 71.2 & 66.8 \\
    & Qwen2-72B & 65.1 & 72.3 & \underline{68.5} \\
    & DeepSeek-V3 & 62.4 & 75.3 & 68.2 \\
    & GPT-4o-mini & 61.7 & 74.3 & 67.4 \\
 \midrule
\multirow{5}{*}{\shortstack{Handcrafted\\\\Few-shot}} 
    & Gemma2-9B & \textbf{\underline{66.2}} & 71.3 & 68.7 \\
    & Llama3.1-8B & 62.1 & 73.6 & 67.4 \\
    & Qwen2.5-7B & 62.1 & 69.6 & 65.6 \\
    & Qwen2.5-32B & 61.2 & \underline{80.4} & \textbf{\underline{69.5}} \\
    & Qwen2-72B & 64.0 & 74.9 & 69.0 \\
    & DeepSeek-V3 & 63.5 & 75.2 & 68.9 \\
    & GPT-4o-mini & 62.3 & 74.0 & 67.6 \\
 \midrule
\multirow{5}{*}{Multipersona} 
    & Gemma2-9B & 56.9  & 71.9 & 63.5 \\
    & Llama3.1-8B & 58.6 & 74.2 & 65.5 \\
    & Qwen2.5-7B & 59.4 & 71.1 & 64.7 \\
    & Qwen2.5-32B & 58.4 & 72.8 & 64.8 \\
    & Qwen2-72B & \underline{61.0} & 75.9 & 67.6 \\
    & DeepSeek-V3 & 60.7 & 77.1 & \underline{67.9} \\
    & GPT-4o-mini & 57.6 & \underline{77.7} & 66.2 \\
\bottomrule
\end{tabular}
\end{table}

% \subsubsection{What is the impact of Composite Prompting Strategies on the MER-OV task?}
% \subsubsection{What impact do Composite Prompting Strategies have on MER-OV performance?\jh{To what extent do composite prompting strategies affect MER-OV performance?}}
\subsubsection{To what extent do composite prompting strategies affect MER-OV performance?}
% \textbf{\textit{Description: }}In the domain of hard prompting for LLMs, several techniques enhance the model’s reasoning capabilities through structured procedural mechanisms, rather than relying on single-step prompts. These techniques can be broadly categorized into three types: Ensembling, Self-Criticism, and Decomposition, which we collectively refer to as Composite Prompting Strategies. Unlike the experiments in the previous section, this part focuses on the further extension of these techniques.
\textbf{\textit{Description: }}We investigated the impact of three composite prompting strategies on the performance of various LLMs on the target task. These composite prompting strategies were used in Stage 2 of the emotion clues-based method.
% To streamline the experimental complexity, we selected one commonly used representative method from each strategy category:

% \begin{itemize}
%     \item \textbf{Ensembling}: Universal Self-Consistency
%     \item \textbf{Self-Criticism}: Self-Refine
%     \item \textbf{Decomposition}: Least-to-Most
% \end{itemize}

\textbf{\textit{Answer: }}\textbf{Universal Self-Consistency and Least-to-Most effectively enhance performance across most LLMs, with Universal Self-Consistency's ensembling strategy yielding the best results. In contrast, Self-Refine is shown to be unstable. Its iterative process, designed for complex reasoning, is inefficient for this task and often leads to erroneous sentiment corrections and performance degradation.}
% Both Universal Self-Consistency and Least-to-Most methods demonstrate improvements across nearly all LLMs tested, with the best performance observed in the ensembling approach represented by Universal Self-Consistency. However, the Self-Refine method appears to be constrained by the task's inherent lack of need for complex logical reasoning, often leading to the overall sentiment being steered in the opposite direction during the iterative process, resulting in unstable effects.

\begin{figure*}[ht]
  \centering 
  \includegraphics[width=1\textwidth]{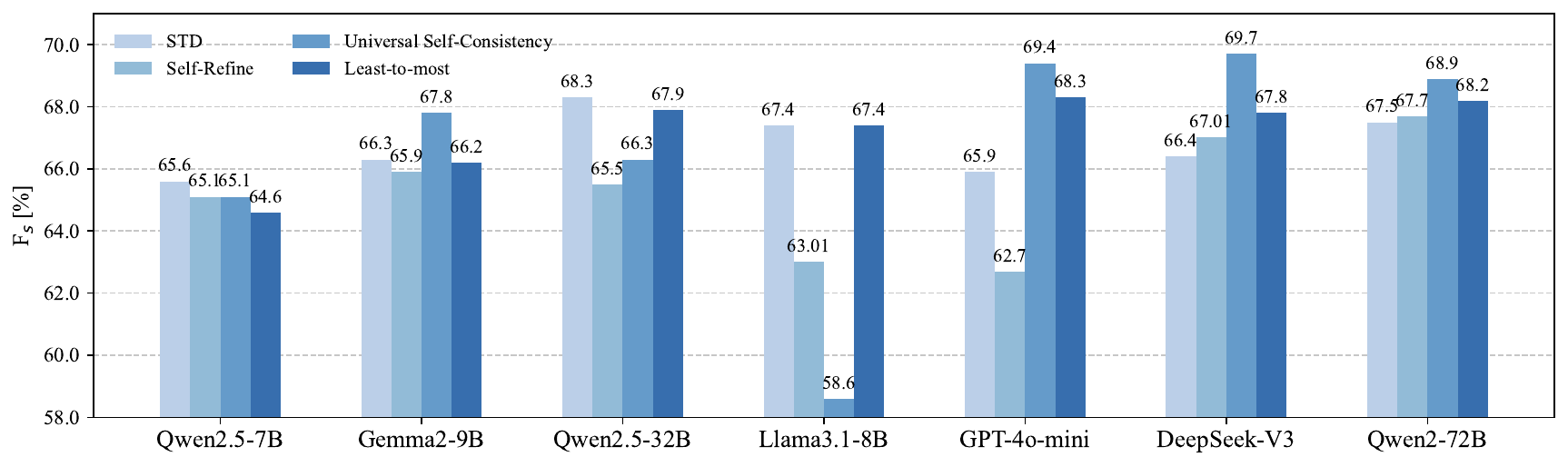} 
  % \caption{Performance of different Composite Prompting Strategies vs. STD on LLMs}
  \caption{Performance evaluation of three selected \textit{composite} (\ie universal self-consistency, self-refine, and least-to-most) and standard (STD) prompting strategies on seven selected LLMs with the emotional clue-based methods. Audio-LLM: Qwen2-Audio-7B; Video-LLM: GPT-4o-mini.}  
  %\textcolor{cyan}{Method: emotional clue-based method. Audio-LLM: Qwen2-Audio-7B; Video-LLM: GPT-4o-mini; LLMs: six open/closed-source LLMs (three on-premises and three API requests). In the second stage, one hard prompt design (standard prompting) and four prompt-based optimization strategies.}}
  \label{Figure 9}
\end{figure*}

\textbf{\textit{Details: }}Figure \ref{Figure 9} presents a comparison of the 
% average 
% (Avg) 
$\text{F}_\text{s}$
%BS: I wrote average (Avg) instead of Avg
scores between three different composite prompting strategies and the STD method across various LLMs.
% For more detailed data metrics, please refer to Table \ref{table:prompt2} in Appendix \ref{Appendix:data-table}. 

From the LLM-centric perspective, the performances of Qwen2.5-7B and Llama3.1-8B are particularly notable. For Qwen2.5-7B, applying different prompting strategies leads to a decrease in Avg performance, primarily due to the model's tendency to provide fewer but more accurate emotional vocabulary, which %BS: significantly --> 
considerably 
lowers $\text{Recall}_\text{s}$ and $\text{Precision}_\text{s}$, consequently, affects overall performance. On the other hand, Llama3.1-8B tends to provide more nuanced and abundant emotional vocabulary, with the Least-to-Most strategy enhancing this tendency. However, the other two strategies do not adapt well, resulting in decreased overall performance. For the other four LLMs, the performance is relatively balanced, with most showing improvements or remaining stable, except for the Least-to-Most strategy on GPT-4o-mini (which also exhibits more detailed but abundant emotional analysis). The best performance is achieved by the combination of Universal Self-Consistency and DeepSeek-V3.

From the perspective of different strategies, the Least-to-Most strategy primarily improves performance by guiding the LLM to generate more detailed sentiment analysis and a richer vocabulary list. In contrast, the Universal Self-Consistency strategy enhances performance by guiding the LLM to select more accurate options from the candidate answers. These two effects have been confirmed across all models, except for Qwen2.5-7B and Llama3.1-8B. Additionally, the performance of the Self-Refine strategy is relatively unstable and lacks a clear pattern. We attribute this to the potential for individual details to negate the original answer during the iterative refinement process, which may shift the overall sentiment in the opposite direction.

\subsubsection{Performance of Reasoning Models in MER-OV.}

% \begin{flushleft}

\textbf{\textit{Description: }}
% Recent advancements in LLMs have prompted a paradigm shift from complex prompt engineering towards leveraging the intrinsic reasoning capabilities of state-of-the-art models. 
% Reflecting this trend, 
We evaluated two prominent Reasoning Models: OpenAI's o3-mini and DeepSeek-R1. To isolate their inherent abilities and ensure a direct comparison against our baseline, we employed the simple prompt configuration from our STD group for these experiments.
% Although we have experimented with various prompting techniques in previous studies, these methods now appear somewhat outdated compared to the rapid advancements in the field of LLMs. Most SOTA LLMs have developed their own deep reasoning models. In this section, we selected OpenAI o3-mini and DeepSeek-R1 as experimental subjects to evaluate their performance on this task. For a more intuitive comparison, the prompt configuration in this section was kept consistent with the STD group setup.

\textbf{\textit{Answer: }}\textbf{Compared to the prompting techniques previously involved in the experiments, deep reasoning LLMs do not fully demonstrate their advantages.}

% On one hand, the task does not impose high demands on complex reasoning. On the other hand, deep reasoning models tend to explore multiple perspectives, with more divergent and creative thinking. While this approach has value in certain contexts, it is not particularly conducive to completing this task.

\textbf{\textit{Details: }}The results are presented in Table \ref{table:prompt4}. We compared Reasoning Models (OpenAI's o3-mini, DeepSeek-R1) against their respective base counterparts (GPT-4o-mini, DeepSeek-V3), yielding two key findings.

First, contrary to expectations, the Reasoning Models offer no significant performance advantage; DeepSeek-R1's average score, in fact, slightly decreases. Second, a distinct shift in prediction strategy was observed. These models achieve a notable balance between $\text{Precision}_\text{s}$ and $\text{Recall}_\text{s}$, whereas baseline models consistently favor high $\text{Recall}_\text{s}$ at the expense of $\text{Precision}_\text{s}$. This suggests the baseline models' high scores may stem from generating a diverse, yet less precise
% , array of emotional labels—a "shotgun" approach.
\begin{table}[!t]
\centering
% \caption{Performance of LLMs with Advanced Reasoning Capabilities in MER-OV task. \textcolor{cyan}{Method: emotional clue-based method. Audio-LLM: Qwen2-Audio-7B; Video-LLM: GPT-4o-mini; LLMs: two open/closed-source LLMs (one on-premises and one API requests) and two Reasoning LLMs (two API request). Question: Should we include Reasoning LLM in Table \ref{table 1}?}}
%
%BS: In the table, one does not understand which is w/ and which is w/o - perhaps add nother column for that.
%
\caption{Performance comparison between LLMs w/ and w/o \textit{advanced reasoning strategies} with the emotional clue-based methods. Audio-LLM: Qwen2-Audio-7B; Video-LLM: GPT-4o-mini.
% ; LLM w/o reasoning: GPT-4o-mini and DeepSeep-V3; LLM w/ reasoning: OpenAI o3-mini and DeepSeek-R1.
% \textcolor{cyan}{Method: emotional clue-based method. Audio-LLM: Qwen2-Audio-7B; Video-LLM: GPT-4o-mini; LLMs: two open/closed-source LLMs (one on-premises and one API requests) and two Reasoning LLMs (two API request). Question: Should we include Reasoning LLM in Table \ref{table 1}?}
}
\label{table:prompt4}
\begin{tabular}{ccccc}

\toprule
Reasoning & LLM & $\text{Precision}_\text{s}$ [\%] & $\text{Recall}_\text{s}$ [\%] & $\text{F}_\text{s}$ [\%] \\
\midrule
\multirow{2}{*}{w/o}
& GPT-4o-mini    & 58.8 & 75.0 & 65.9 \\
& DeepSeek-V3    & 59.9  & 74.6 & \textbf{66.4} \\
\midrule
\multirow{2}{*}{w/}
& OpenAI o3-mini & 65.9 & 68.5 & \textbf{67.2} \\
& DeepSeek-R1    & 64.4  & 68.9 & 66.6 \\
\bottomrule
\end{tabular}
\end{table}
An analysis of DeepSeek-R1 elucidates these findings. The model enhances its base version (DeepSeek-V3) by integrating a reasoning content generation process based on reinforcement learning that analyzes implied contexts and potential emotional transitions. However, this creates a fundamental capability-task mismatch. The model's sophisticated, nuanced reasoning is ``over-engineered'' for a dataset demanding direct, short-term emotion identification. This over-analysis explains both the lack of a performance increase and the shift towards a more balanced, deliberate $\text{Precision}_\text{s}$-$\text{Recall}_\text{s}$ profile, as the model attempts to pinpoint a single, deeply reasoned (often overly complex) emotional state rather than listing all possibilities.

\section{Conclusion}
\label{sec:conclusion}
In this work, we presented a comprehensive and systematic investigation into Open-Vocabulary Multimodal Emotion Recognition (MER-OV), a novel paradigm that advances emotion understanding in the era of large models. This extensive evaluation spans foundational frameworks, modality contributions, model architectures, and prompting strategies, establishing some of the first crucial benchmarks in this nascent field. By conducting large-scale experiments across leading open-source and closed-source LLMs, Video-LLMs, and Audio-LLMs, we demonstrate the efficacy of a two-stage method that first extracts descriptive emotional clues from each modality before final inference, with trimodal fusion of audio, video, and text proving the most effective combination and video emerging as the most critical contributor. Our analysis further reveals that the performance gap between leading open-source and closed-source LLMs is surprisingly narrow;
% with smaller models like Llama3.1-8B performing exceptionally well; 
in contrast, closed-source Video-LLMs hold a significant advantage that scales positively with model size, while differences among Audio-LLMs are less pronounced. Efforts to enhance video input through dynamic frame sampling or metadata inclusion yielded no substantial improvements, indicating that the current bottleneck resides in the intrinsic understanding capabilities of Video-LLMs rather than input richness. Additionally, task-specific handcrafted prompts outperform generic ones, and advanced prompt engineering techniques provide consistent gains, though specialized Reasoning Models show no clear advantage in this context. 
Looking ahead, the establishment of a robust MER-OV paradigm hinges on progress along three key directions: (1) the creation of new, dedicated datasets to enable comprehensive benchmarking and stress-test model generalizability; (2) rigorous multilingual and multicultural evaluations to assess robustness across diverse linguistic and social contexts; and (3) the integration of advanced multimodal fusion techniques and state-of-the-art vision-language models. By advancing these initiatives, we can pave the way for a new generation of empathetic AI systems, which will revolutionize real-world human-computer interaction by enabling more natural and responsive applications in areas like mental health support and personalized companion agents.

\bibliographystyle{IEEEtran}
\bibliography{Reference}

\vfill

\end{document}